\documentclass{SciPost}

\hypersetup{
    colorlinks,
    linkcolor={red!50!black},
    citecolor={blue!50!black},
    urlcolor={blue!80!black}
}

\usepackage[bitstream-charter]{mathdesign}
\DeclareSymbolFont{usualmathcal}{OMS}{cmsy}{m}{n}
\DeclareSymbolFontAlphabet{\mathcal}{usualmathcal}

\fancypagestyle{SPstyle}{
\fancyhf{}
\lhead{\colorbox{scipostblue}{\bf \color{white} ~SciPost Physics }}
\rhead{{\bf \color{scipostdeepblue} ~Submission }}

\fancyfoot[C]{\textbf{\thepage}}
}

\usepackage{bm} 
\usepackage{empheq}
\usepackage{amsthm}
\newtheorem{proposition}{Proposition}
\usepackage{caption}
\usepackage{tabularx}
\usepackage{booktabs}
\usepackage{float}
\usepackage{wrapfig}   

\usepackage{booktabs}

\begin{document}

\pagestyle{SPstyle}

\begin{center}{\Large \textbf{\color{scipostdeepblue}{
Kaleidoscope Yang-Baxter Equation for Gaudin's Kaleidoscope models\\
}}}\end{center}

\begin{center}\textbf{
Wen-Jie Qiu \textsuperscript{1,2},
Xi-Wen Guan \textsuperscript{1, 3, 4} and
Yi-Cong Yu \textsuperscript{1$\dagger$}
}
\end{center}

\begin{center}
{\bf 1} Innovation Academy for Precision Measurement Science and Technology, Chinese Academy of Sciences, Wuhan 430071, China\\
{\bf 2} University of Chinese Academy of Sciences, Beijing 100049, China \\
{\bf 3} Hefei National Laboratory, Hefei 230088, People’s Republic of China \\
{\bf 4} Department of Fundamental and Theoretical Physics, Research School of Physics, Australian National University, Canberra, ACT 0200, Australia\\

$\dagger$ \href{mailto:}{\small ycyu@wipm.ac.cn}
\end{center}

\section*{\color{scipostdeepblue}{Abstract}}
\textbf{
\boldmath{
%
Recently, researchers have proposed the Asymmetric Bethe ansatz method-a theoretical tool that extends the scope of Bethe ansatz-solvable models by ``breaking" partial mirror symmetry  via the introduction of a fully reflecting boundary. 
Within this framework, the integrability conditions which were originally put forward by Gaudin have been further generalized.
In this work, building on Gaudin's generalized kaleidoscope model,  we present a detailed investigation of  the relationship between $D_{N}$ symmetry and its integrability.
We demonstrate that the mathematical essence of integrability in this class of models is characterized by a newly proposed Kaleidoscope Yang-Baxter Equation.
Furthermore, we show that the solvability of a model via the coordinate Bethe ansatz depends not only on the consistency relations satisfied by scattering matrices, but also on the model’s boundary conditions and the symmetry of the subspace where solutions are sought.
Through finite element method (based numerical studies), we further confirm that Bethe ansatz integrability arises in a specific symmetry sector.
Finally, by analyzing the algebraic structure of the Kaleidoscope Yang–Baxter Equation, we derive a series of novel quantum algebraic identities within the framework of quantum torus algebra.
}
}

\vspace{\baselineskip}

\noindent\textcolor{white!90!black}{%
\fbox{\parbox{0.975\linewidth}{%
\textcolor{white!40!black}{\begin{tabular}{lr}%
  \begin{minipage}{0.6\textwidth}%
    {\small Copyright attribution to authors. \newline
    This work is a submission to SciPost Physics Core. \newline
    License information to appear upon publication. \newline
    Publication information to appear upon publication.}
  \end{minipage} & \begin{minipage}{0.4\textwidth}
    {\small Received Date \newline Accepted Date \newline Published Date}%
  \end{minipage}
\end{tabular}}
}}
}


\vspace{10pt}
\noindent\rule{\textwidth}{1pt}
\tableofcontents
\noindent\rule{\textwidth}{1pt}
\vspace{10pt}


\section{Introduction}
\label{sec:intro}
The coordinate Bethe ansatz (CBA) was first introduced by Hans Bethe in 1931 in his study of the one-dimensional Heisenberg spin chain model \cite{bethe1931}.
This groundbreaking work established the theory of quantum integrable systems.
Bethe postulated that the wave function of a one-dimensional many-body system-within each region of fixed particle ordering can be expressed as a linear combination of finitely many plane waves.
Coefficients across distinct regions are linked by constraints from particle interactions and boundary conditions.
These constraints reduce to a set of transcendental algebraic equations for quasi-momenta, termed the Bethe equations.
In subsequent developments, the CBA has been extended to a range of prominent one-dimensional systems: the Lieb-Liniger model \cite{lieb1963,lieb1963a} describing 1D bosonic gases with delta-function interactions, the Yang-Gaudin model \cite{yang1967a} (spinful Fermi gases), the XXZ \cite{yang1966c} and XYZ \cite{baxter1973a, baxter1973} spin chains, and the Hubbard model \cite{lieb1968a}. 
This expansion has greatly broadened the scope of the CBA  into one-dimensional integrable systems.

The CBA’s core significance lies in its provision of a complete, exact solution framework for 1D quantum many-body systems, enabling the rigorous calculation and analysis of key physical properties, including energy spectra, ground/excited states, and correlation functions of such systems \cite{takahashi1999, caux2011}.
This approach has not only underpinned the development of 1D quantum many-body theory but also driven progress in cutting-edge fields: condensed matter physics \cite{takigawa1996, lake2005, tennant1995}, quantum statistics \cite{baxter1985}, and cold atom physics \cite{cazalilla2011, guan2013a}. 
Its theoretical predictions have also shown excellent agreement with experimental results \cite{kinoshita2004, paredes2004, pagano2014, guan2022a}.
Furthermore, the CBA is tightly coupled to the mathematical structures inherent to integrable models including group theory, symmetries, and quantum groups \cite{drinfeld1985, jimbo1985}, establishing it as an indispensable theoretical tool for investigating integrable systems and quantum many-body problems.

For the Lieb-Liniger model \cite{lieb1963,lieb1963a}, M. Gaudin developed a concrete, systematic solution framework (see Chapter 5 of \cite{gaudin2014}).
Within this framework, Gaudin solved mirror systems via the Bethe ansatz.
These mirror systems exhibit a generalized kaleidoscope structure—composed of $\delta$-function mirrors and invariant under reflection across each mirror.
In the absence of coupling constants, such mirror systems are classified by finite reflection groups \cite{humphreys1992}.
A necessary (but not sufficient) condition for reflection invariance of a mirror set is that dihedral angles between mirrors take the form 
$\pi/n$, where $n$ is a positive integer. 
Building on Gaudin’s framework, it was pointed out in  \cite{jackson2024}  that coupling constants of any two mirrors intersecting at $\pi/n$ must be equal.

In this context, a notable example is the solution to the Liu-Qi-Zhang-Chen (LQZC) model \cite{liu2019}—a system once thought to violate the necessary integrability conditions \cite{jackson2024}.
Building on Gaudin’s framework, M. Olshanii and collaborators recently introduced the Asymmetric Bethe ansatz (ABA) method \cite{olshanii2025, jackson2024}.
This approach lifts prior symmetry-related restrictions on sufficient integrability conditions, extending the Bethe ansatz to a wider class of models with 
$\delta$-function interaction potentials.

In this work, we present a detailed analysis of Gaudin’s generalized kaleidoscope framework in a two-dimensional system.
We demonstrate that the solvability of the model via the coordinate Bethe ansatz (CBA) is equivalent to the existence (over the solution domain) of a Bethe ansatz, defining  vector bundle section.
A necessary condition for this existence of section can  be cast in terms of the Kaleidoscope Yang-Baxter Equation (KYBE).
We also establish a systematic approach to deriving Bethe ansatz equations via concrete examples and show that a system’s CBA solvability depends not only on the consistency relations satisfied by scattering matrices, but also on the model’s boundary conditions and the symmetry of the subspace where solutions are sought.
Unlike Gaudin’s original treatment, which adopted a trivial symmetry group representation for simplicity, our work generalizes Gaudin’s framework and uncovers the intricate relationship between integrability and the symmetry group.
Furthermore, we note that the KYBE exhibits a nontrivial mathematical structure and is deeply linked to quantum torus algebra.

\section{Coordinate Bethe ansatz in two-dimensional plane \label{sec2}}
\begin{figure}[htbp]       
  \centering               %
  \includegraphics[width=0.5\textwidth]{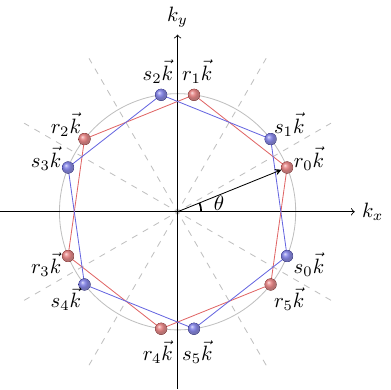} 
  \caption{A schematic of the $D_6$ group. Here, ${}_{ r_m}\vec{k} $ and ${}_{s_n}\vec{k}$ denote the points obtained by applying rotation ${ r_m}$  or reflection $ s_n$ operations, respectively, to the initial object (for example, the momentum vector 
$\vec{k}$, which corresponds to the point $r_0 := \mathrm{id}$), $\theta$ denotes the initial angle. }
  \label{figDN}       %
\end{figure}

The coordinate Bethe ansatz is an analytical method in which the many-body wave function is expressed, within each distinct region of solution domain, as a linear combination of plane waves, with the coefficients determined by scattering and boundary conditions \cite{gaudin2014}.
We focus on the two-dimensional case and consider particle scattering in a potential that is invariant under the $D_{N}$ group \cite{liu2019}. In this case, the wave function can be generally expressed as
\begin{equation}
    \Psi(\vec{x}) =\sum_{g \in D_{N}} A_g\mathrm{e}^{\mathrm{i}(\vec{x},g\vec{k})}, \quad \vec{x} \in D_F.
    \label{eqBA}
\end{equation}
The $D_F$ denotes the fundamental region, which is a convex subset within the solution domain. 
In the above equation, $(\vec{x},g\vec{k})$ denotes the inner product. 
Under reflection group action, this region generates a family of convex subsets whose union constitutes the full solution domain, and the intersection of any two such subsets has measure zero  (see \cite{gaudin2014}, Chapter 5, formula 5.28). 
In simple cases, the Bethe ansatz wave function is a pure superposition of plane waves within $D_F$,  while the wave function in the remaining solution domains follows from symmetry group operations on $D_F$.
$\vec{k}$ is the Bethe root, $A_g$ is the amplitude, and $D_{N}$ denotes the dihedral group with $2N$ elements, in which the group elements satisfy:
\begin{align}
    r_mr_n=r_{m+n},\quad r_ms_n=s_{m+n}, \quad
    s_mr_n=s_{m-n},\quad s_ms_n=r_{m-n}
    \label{eqDN}
\end{align}
for $m, n = 0, 1, \cdots,N-1$. 
The addition and subtraction in (\ref{eqDN}) are both performed modulo 
$N$. Fig.\ref{figDN} provides a schematic of group $D_6$.
The initial momentum $k$ after being acted upon by elements of the group 
$g$ is uniformly distributed along the circumference of a circle. 
 As a group element, $r_m$ represents a rotation about the origin by an angle of $2\pi m/N$, while $s_n$ represents a reflection with respect to the axis at an angle of $\pi/N$. 
 In the subsequent  discussion, we use $s_k$ to denote a barrier, i.e. a reflection with respect to this barrier,  corresponding to the symmetry group operation 
$s_k$. 

The form of the Bethe ansatz wave function in Equation (\ref{eqBA}) naturally introduces a section of a vector bundle over the entire solution domain. 
If we regard $A_g$ as a vector of dimension $2N$, then at each point in the solution domain we attach a vector.
 In the fundamental region $D_F$,  this vector bundle can be defined as
\begin{align}
    A_g(\vec{x})=A_g\cdot\mathrm{e}^{\mathrm{i}(\vec{x}, g\vec{k})}.
    \label{eqVB}
\end{align}
 and in other regions (coordinate charts), the vector bundle can be determined by scattering, which will be discussed below. 

 Note that the symmetry of the wave function in Equation (\ref{eqBA}) can be reformulated as a symmetry property of the section of the vector bundle. 
 In this regard, we state the following proposition without proof:
 \begin{proposition}\label{thmSymmetry}
Suppose $\Psi$ and $\Psi^\prime$ are two Bethe ansatz wave functions of the form (\ref{eqBA}), while $\bm{A}$ and $\bm{A}^\prime$ correspond to the sections reduced by $\Psi$ and $\Psi^\prime$,  respectively. If $\Psi^\prime$ is obtained from $\Psi$ by a symmetry operation described by $\eta \in D_{N}$, i.e., $\Psi^\prime(\eta^{-1}\vec{x}) = \Psi(\vec{x})$, then at the origin point, we have $\bm{A}^\prime _g = \bm{A}_{\eta^{-1}g}$.
\end{proposition}

The vector bundle induced by the form of the Bethe ansatz wave function possesses the following property: if the attached vector at a single point in the solution domain is known, then the attached vector at any other point in the domain can be determined starting from that point.
 Below, we present the rules that govern the ``evolution'' of these vectors from a known point $A$ to other  points $B$, which can be categorized into three main types:

\begin{enumerate}
    \item {\bf Free propagation}
    
     In this case, points $A$ and $B$ are located within the same region and can be connected by a straight line segment. 
    Then according to (\ref{eqVB}), $B_g=A_g\cdot\mathrm{e}^{\mathrm{i}(\vec{x}_B - \vec{x}_A, g\vec{k})}$. 
    If we denote the vector with  the basis  
    \begin{equation}
    \bm{A}:=[A_{r_0},A_{r_1}, \cdots, A_{r_{N-1}},A_{s_0},A_{s_{-1}},\cdots,A_{s_{-(N-1)}}]^T,
    \end{equation}
   where $T$ stands for the transposition.
    Then the vector at $B$ and $A$ can be linked by a diagonal matrix
    \begin{equation}
    \bm{\beta}_{\vec{\Delta x}}:=\mathrm{diag}(\mathrm{e}^{\mathrm{i}(\vec{\Delta x}, r_0 \vec{k})}, \mathrm{e}^{\mathrm{i}(\vec{\Delta x}, r_1 \vec{k})}, 
    \cdots,
    \mathrm{e}^{\mathrm{i}(\vec{\Delta x}, r_{N-1} \vec{k})},
    \mathrm{e}^{\mathrm{i}(\vec{\Delta x}, s_0 \vec{k})}, \mathrm{e}^{\mathrm{i}(\vec{\Delta x}, s_{-1} \vec{k})},
    \cdots,
    \mathrm{e}^{\mathrm{i}(\vec{\Delta x}, s_{-(N-1)} \vec{k})}
    )
    \label{eqbeta}
\end{equation}
with $\vec{\Delta x}:=\vec{x}_B - \vec{x}_A$. In simple terms, free propagation can be represented by left multiplication with diagonal matrix defined by (\ref{eqbeta})
{\setlength{\fboxrule}{0.2pt}
\begin{empheq}[box=\fbox]{align}
  \bm{A} \to \bm{\beta} \bm{A}.
    \label{eqRule1}
\end{empheq}
}

\begin{figure}[t]       
  \centering               
  \includegraphics[width=0.85\textwidth]{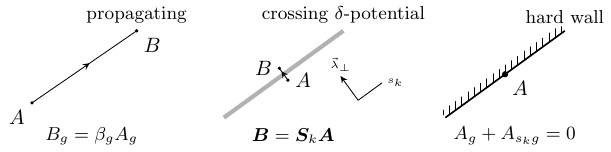} 
  \caption{Schematics of the three types of rules. The left panel represents propagation within the same region, which can be described by a diagonal matrix (\ref{eqbeta}). The middle panel depicts scattering at the potential barrier, which can be represented by a scattering matrix (\ref{eqScat}). The right panel shows the boundary condition at the hard wall, which can be expressed as a constraint (\ref{eqRule3}).}
  \label{figRules}       
\end{figure}

    \item{\bf Crossing $\delta$-potential barriers}
    
    This is a typical coordinate Bethe ansatz problem. 
    The wave function satisfies two conditions \cite{lieb1963, lieb1963a}: (i) continuity; and (ii) the discontinuity in its derivative of the wave function along the direction perpendicular to the barrier equals  the interaction constant $c$ times  the value of the wave function on the barrier.
    Let the reflection across the barrier be denoted  by $s_k$, and the reflection with respect to the axis perpendicular to the barrier by $s_{k\perp}$.
     Since $N$ is even, $s_{k \perp}$ also lies in the  $D_{N}$ group. 
     Along the barrier, the wave function on side $A$ can be  given by 
    \begin{equation}
        \Psi_A = \sum_{g \in D_N}A_g \mathrm{e}^{\mathrm{i}(\vec{x}, g \vec{k})}
        = \sum_{g \in D_N /\{I, s_k\}} (A_g + A_{s_kg}) \mathrm{e}^{\mathrm{i}(\vec{x}, g \vec{k})},
        \label{eqPsiA}
    \end{equation}
    here, $D_N/{\{I,s_k \}}$ denotes the quotient group obtained by dividing $D_N$ by its subgroup $\{I, s_k \}$. 
    The second equality in (\ref{eqPsiA}) holds because $(\vec{x}, g\vec{k}) = (\vec{x}, s_kg\vec{k})$ as $\vec{x}$ locates at the barrier. 
    Similarly, the derivative in the direction perpendicular to the barrier can be expressed as
    \begin{equation}
        \nabla_{\vec{\lambda}_{\perp}} \Psi_A = \sum_g \mathrm{i}(\vec{\lambda}_{\perp}, g\vec{k})A_g
        \mathrm{e}^{\mathrm{i}(\vec{x}, g\vec{k})} 
        = \sum_{g \in D_N /\{I, s_k\}} \mathrm{i}(\vec{\lambda}_{\perp}, g\vec{k})
        \left(A_g - A_{s_kg}\right) \mathrm{e}^{\mathrm{i}(\vec{x}, g \vec{k})},
        \label{eqDerPsiA}
    \end{equation}
    here $\vec{\lambda}_\perp$ denotes the unit vector across the barrier between  $A$ and  $B$, in other words,  perpendicular to the barrier. 
    For the wave function on side $B$ of the barrier, we can obtain expressions entirely analogous to (\ref{eqPsiA}) and (\ref{eqDerPsiA}). Using  the conditions  $\Psi_B = \Psi_A$ and $\nabla_{\vec{\lambda}_\perp}(\Psi_B - \Psi_A) = c\cdot \Psi_A$  which hold  at the barrier, we then  obtain 
    \begin{equation}
        B_g = \left(1+ \frac{c}{2\mathrm{i}(\vec{\lambda}_\perp,g\vec{k})}\right) A_g + \frac{c}{2\mathrm{i}(\vec{\lambda}_\perp,g\vec{k})}A_{s_kg}.
        \label{eqScat}
    \end{equation}
    It should be noted that in (\ref{eqScat}), the dependence of the coefficients on the symmetry group element $s_k$ is reflected in the vector $\vec{\lambda}_\perp$. 
    Here, 
$\vec{\lambda}_\perp$ is the unit vector perpendicular to the symmetry axis, as illustrated in the middle panel of Figure \ref{figRules}. 
For symmetry axis $s_k$, the form of $\vec{\lambda}_\perp$ is given by $[\cos(\frac{\pi k}{N} + \frac{\pi}{2}), \sin(\frac{\pi k}{N} + \frac{\pi}{2})]$, and the coefficients in (\ref{eqScat}) can be computed explicitly.

    We rearrange the result of Equation (\ref{eqScat}), introducing $N \times N$  Toeplitz matrix $\bm{t}$ and diagonal matrix $\bm{s}$ as follows:
    \begin{equation}
        \bm{t} :={
        \begin{bmatrix}
            0 & 1 & 0 & \cdots & 0 \\
            0 & 0 & 1 & \cdots & 0 \\
            0 & 0 & 0 & \cdots & 0 \\
            \vdots & \vdots & \vdots & \ddots & \vdots \\
            1 & 0 & 0 & \cdots & 0
        \end{bmatrix}_{N \times N}}
        \bm{s} :={
        \begin{bmatrix}
            (\sin \theta)^{-1} & 0  & \cdots & 0 \\
            0 & [\sin (\frac{2\pi}{N} + \theta)]^{-1}  & \cdots & 0 \\
            \vdots & \vdots & \ddots & \vdots \\
            0 & 0 & \cdots & [\sin (\frac{2\pi(N-1)}{N} + \theta)]^{-1}
        \end{bmatrix}}.
        \label{eqts}
    \end{equation}
    In the above expressions, we have reparametrized the Bethe roots in (\ref{eqBA}) in terms of $k_0>0$ with  $\theta$ via  $\vec{k} = [k_0\cos\theta, k_0\sin\theta]$. 
     These parameters help us to write a simple form for the operations of the $D_N$ group elements $r_m$ and the reflection $s_n$ to the Bethe roots $\vec{k}$: 
    \begin{eqnarray}
    r_m\vec{k} &=& [k_0\cos(\frac{2\pi m}{N} + \theta), k_0\sin(\frac{2\pi m}{N} + \theta)],\nonumber\\
     s_n\vec{k} &=& [k_0\cos(\frac{2\pi n}{N} - \theta), k_0\sin(\frac{2\pi n}{N} - \theta)],\nonumber
     \end{eqnarray}
     respectively. 
    Then  using (\ref{eqts}) we construct $2N \times 2N$ block matrices
    \begin{equation}
    \bm{T}:=
\begin{bmatrix}
\bm{t} & \bm{0} \\
\bm{0} & \bm{t}^{-1}  \\
\end{bmatrix}, \quad
\bm{S}_0:=
\begin{bmatrix}
\bm{I}+\frac{c}{2\mathrm{i}k_0}\bm{s} & \frac{c}{2\mathrm{i}k_0}\bm{s}  \\
-\frac{c}{2\mathrm{i}k_0}\bm{s}  & \bm{I}-\frac{c}{2\mathrm{i}k_0}\bm{s}   \\
\end{bmatrix}.
\label{eqTS}
\end{equation}
With the help of the  above definitions, we can readily write down the scattering matrix for the barrier along the $s_k$ axis \footnote{Where $k$ is even, the case with odd $k$ can be described by a similar expression, requiring only a redefinition of the $\bm{s}$-matrix.} 
\begin{equation}
    \bm{S}_k=\bm{T}^{-k/2}\bm{S}_0\bm{T}^{k/2}.
    \label{eqSk}
\end{equation}
Consequently, the "evolution"  across the $\delta$-potential barrier can be written by 
{\setlength{\fboxrule}{0.2pt}
\begin{empheq}[box=\fbox]{align}
  \bm{A} \to \bm{S}_k \bm{A}.
    \label{eqRule2}
\end{empheq}
}
    
    \item{\bf Hard-wall boundary condition}
    
    A hard wall imposes a Dirichlet-type boundary condition, i.e,  the value of the wave function at the hard wall is zero. 
    According to the expression (\ref{eqPsiA}) we obtain $A_g + A_{s_kg} = 0$. 
   Defining 
    \begin{equation}
    \bm{\Gamma}_k=
\begin{bmatrix}
\bm{0} & \bm{t}^{-k} \\
\bm{t}^{k} & \bm{0}  \\
\end{bmatrix},
\end{equation}
then we  can express the  boundary condition in the matrix form:
{\setlength{\fboxrule}{0.2pt}
\begin{empheq}[box=\fbox]{align}
  \bm{A} + \bm{\Gamma}_k \bm{A} = 0.
    \label{eqRule3}
\end{empheq}
}
\end{enumerate} 
The three rules (\ref{eqRule1}), (\ref{eqRule2}), and (\ref{eqRule3}) outlined above fully determine the behavior of vector bundle sections over the solution domain.

\begin{figure}[t]       
  \centering               %
  \includegraphics[width=0.8\textwidth]{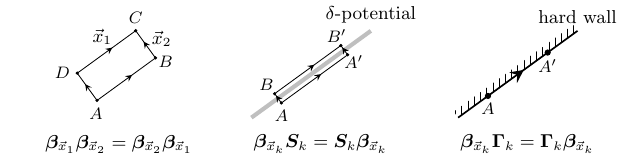} 
  \caption{Schematics of the consistency conditions. The left panel illustrates the equivalence of different paths in the free propagation region, which follows directly from the fact that the matrices $\bm{\beta}_{\vec{x}}$ are diagonal. The middle panel depicts the equivalence between two different paths starting from point A: one path crosses the potential barrier first and then propagates along the barrier, while the other path first propagates along the barrier and then crosses the barrier. This equivalence is also manifested as the commutativity of the matrices $\bm{\beta}_{\vec{x}_k}$ and $\bm{S}_k$. The right panel represents the consistency of the boundary conditions, i.e., a point A that satisfies the boundary condition will continue to satisfy the boundary condition after propagation along the hard wall direction. This property can likewise be established through the commutativity of the matrices $\bm{\beta}_{\vec{x}_k}$ and $\bm{\Gamma}_k$.}
  \label{figCons}       %
\end{figure}

These rules must satisfy consistency requirements: whenever two points can be connected by different paths in the solution domain, the resulting amplitude transport must be path independent.
These conditions are a core prerequisite for integrability \cite{yang1967a, baxter1985} and are straightforward to verify (see Figure \ref{figCons}).
For paths confined to the same region, the relevant $\bm{\beta}$ matrices are diagonal and commute, so consistency holds trivially. 
A nontrivial case occurs when both paths cross a  $\delta$-potential barrier (see middle panel of Figure \ref{figCons}), where consistency relies on commutativity between 
matrices $\bm{\beta}_{\vec{x}_k}$ and $\bm{S}_k$. 
 Here $\vec{x}_k = [r_0\cos\frac{\pi k}{N}, r_0\sin\frac{\pi k}{N}]$ denotes the displacement along the direction of the potential barrier. 
 Without losing generality, we only prove the $k=0$ case,  $\bm{\beta}_{\vec{x}_k}$ for different barrier directions obey transformation relations analogous to Eq. (\ref{eqSk}). 
 Commutativity $[\bm{\beta}_{r_0\vec{e}_x}, \bm{S}_0] = 0$   can be  verified with the help of  the equations (\ref{eqbeta}), (\ref{eqts}), and (\ref{eqTS}).
 Additionally, boundary condition consistency requires $(\bm{I} + \bm{\Gamma}_k) \bm{A} = 0$ , which implies  $(\bm{I} + \bm{\Gamma}_k) \bm{\beta}_{\vec{x}_k}\bm{A} = 0$. 
This also follows  directly from $[\bm{\beta}_{\vec{x}_k}, \bm{\Gamma}_k]=0$.

In the preceding discussion, we establish the rules governing sections of the vector bundle.
This vector bundle is directly linked to the Bethe ansatz—thus, finding coordinate Bethe ansatz solutions reduces to identifying self-consistent vector bundle sections that satisfy these rules. 
We will present  the merits of this framework: for additional physical terms in the Hamiltonian, the vector bundle formalism naturally encodes multi-scattering consistency into the multicomponent plane-wave ansatz. 
The non-trivial holonomy of the bundle enforces the Kaleidoscope Yang-Baxter Equation (KYBE), which we elaborate on in subsequent sections.
This approach extends to higher dimensions, though this generalization is not pursued in the present work.

\begin{figure}[t]       
  \centering               %
  \includegraphics[width=0.8\textwidth]{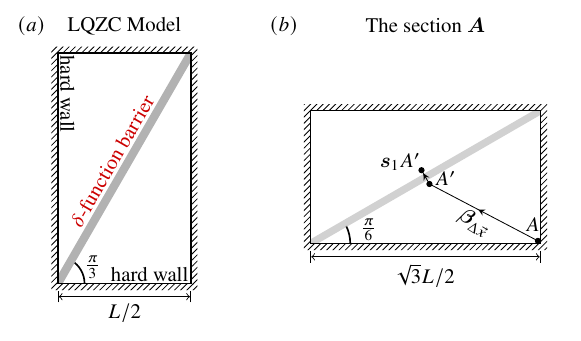} 
  \caption{(a) 
   Schematic phase diagram of the Liu-Qi-Zhang-Chen model \cite{jackson2024}.
The original 1D two-body problem is equivalent to a 2D single-particle problem, i.e. a quantum particle confined in a box of size   $\frac{1}{2}L \times \frac{\sqrt{3}}{2}L$. 
(b) Consider the section $A$  of the vector bundle over this region.
For computational convenience, we rotated the model clockwise by  $90^\circ$.  }
  \label{figLQZC}       %
\end{figure}

\section{Liu-Qi-Zhang-Chen model revisited}
\label{sec:another}

Liu, Qi, Zhang, and Chen (LQZC) \cite{liu2019} first solved the quantum mechanical problem of two particles with a mass ratio of $3:1$
 confined in a one-dimensional hard-wall potential.
They derived an explicit solution via the coordinate Bethe ansatz.
Notably, this solution violates Gaudin’s necessary conditions for conventional integrability \cite{gaudin2014, jackson2024}.
It can be interpreted as a generalization of the Lieb-Liniger model \cite{lieb1963, lieb1963a}.
The model schematic is shown in Figure \ref{figLQZC}(a).
Taking the rectangle’s lower-left corner as the origin, we express the Hamiltonian as:
\begin{equation}
    \hat{H}=-\frac{\partial^2}{\partial x^2} - \frac{\partial^2}{\partial y^2} + c \delta(\frac{\sqrt{3}}{2}x-\frac{1}{2}y),
    \label{eqLQZC}
\end{equation}
where $c$ is the coupling constant. The LQZC model was solved by Bethe ansatz of the form (\ref{eqBA}) with $N = 6$ \cite{liu2019}. 

Here, we revisit this model and develop a systematic method for solving such systems.
Our primary result is a general conclusion on the Bethe ansatz solution: for the  $D_{N}$  symmetry group with  $N \ge 4$, the Bethe ansatz equations for the system of Eq.(\ref{eqLQZC}) are generically overconstrained, meaning the system cannot be solved via the conventional coordinate Bethe ansatz.
As a complementary result, we numerically verify that for the special $D_6$  case, an explicit set of quantum numbers generates a complete family of Bethe-ansatz states and reproduces the full spectrum.

\subsection{The Bethe ansatz equations}

We consider the section $\bm{A}$ over the solution domain, as being shown in Figure \ref{figLQZC} (b). 
Here we have rotated the original model clockwise by $90^\circ$.  
this is done to align the symmetry axis corresponding to the $\delta$-function barrier with $s_1$, which simplifies our calculations. (In fact, for any dihedral group $D_N$ with $N$ even, for computational convenience, we require that the $\delta$-function barrier corresponds to a symmetry axis $s_k$ with $k$ being odd. It is straightforward to show that this can always be achieved.

We  first note that  this model possesses a $\mathbb{Z}_2$-type symmetry corresponding to a $\pi$-rotation, which allows us to decompose the solution space into sectors of even and odd parity.
 We consider the sections on either side of the center of this rotation, as shown in Figure \ref{figLQZC}(b); their relationship can be described by the rules (\ref{eqRule2}) we established earlier.
  Therefore these two sections can be denoted as $\bm{A}^\prime$ and $\bm{S}_k\bm{A}$. Moreover, due to the $\pi$-rotational symmetry and in light of Proposition \ref{thmSymmetry} stated previously, we can build the equation $\bm{S}_k\bm{A}^\prime = \pm\bm{T}^{N/2} \bm{A}^\prime$. Here $\bm{T}$ is the rotation matrix defined in (\ref{eqTS}), and the plus and minus signs correspond to even and odd parity, respectively. Additionally, note that $\bm{A}$ at the lower right corner simultaneously satisfies two boundary conditions described by rule (\ref{eqRule3}), and its relation to $\bm{A}^\prime$ is described by rule (\ref{eqRule1}). Thus, we obtain the following equations for scattering process:
\begin{equation}
        \begin{cases}
            \bm{T}^{N/2}\bm{A}^\prime=\pm\bm{S}_k\bm{A}^\prime, \\
            \bm{A}^\prime=\bm{\beta}_{\vec{\Delta x}} \bm{A} ,\\
            \bm{\Gamma}_0\bm{A}+\bm{A}=0, \\
            \bm{\Gamma}_{N/2}\bm{A}+\bm{A}=0,
        \end{cases}
        \label{eqBAA}
\end{equation}
here $\vec{\Delta x}$ denotes the displacement from $\bm{A}$ to $\bm{A}^\prime$. 
Next, using Equations (\ref{eqBAA}), we derive the Bethe ansatz equations for the model with  an even number of particles  $N$ and odd $k$.

For simplicity, we use $\theta_g$ to denote the coefficient in the scattering equation (\ref{eqScat}), where $\theta_g=\frac{c}{2\mathrm{i} (\vec{\lambda}_{k\perp}, g\vec{k})}$. 
Then by considering the first two equations in the system (\ref{eqBAA}), we obtain the following equation:
\begin{equation}
    (1+\theta_g)\beta_gA_g + \theta_g\beta_{s_kg}A_{s_kg} = \pm \beta_{r_\frac{N}{2}}A_{r_\frac{N}{2}g}.
     \label{eqBAAC}
\end{equation}
Substituting $g=r_m$ into (\ref{eqBAAC}), applying the group operation (\ref{eqDN}), and taking into account the boundary conditions represented by the last two equations in (\ref{eqBAA}), we can obtain
\begin{equation}
    \left( (1 + \theta_{r_m})\beta_{r_m} \mp \beta_{r_{\frac{N}{2}+m}}  \right) A_{r_m} = \theta_{r_m}\beta_{s_{k-m}}A_{r_{m-k}}.
    \label{eqBAAT}
\end{equation}
Equation (\ref{eqBAAT}) provides a recursive form for the components of $\bm{A}$, which we express as follows:
\begin{equation}
    A_{r_{m-k}} = f_m A_{r_m}, \quad f_m:=\frac{(1+\theta_{r_m})\beta_{r_m} \mp \beta_{r_m}^{-1}}{\theta_{r_m}\beta_{s_{k-m}}}.
    \label{eqBAAT2}
\end{equation}
The boundary conditions give, $A_{r_{m + N/2}} = A_{r_m}$. Moreover, since we have assumed that $k$ is odd and $N$ is even, it follows that $m - (N/2) \cdot k \equiv m - N/2 \pmod{N}$. Then we can derive 
\begin{equation}
    A_{r_m} = A_{r_{m-\frac{N}{2}}} = A_{r_{m - \frac{N}{2} k}}=f_mf_{m-k}\cdots f_{m-(\frac{N}{2}-1)k}A_{r_m}.
\end{equation}
It follows by the self-consistency condition
\begin{equation}
    f_mf_{m-k}\cdots f_{m-(\frac{N}{2}-1)k} = 1, \quad \mathrm{for} \,\, m = 0, 1, \cdots, (N-1). 
    \label{eqConsistent}
\end{equation}
It is also straightforward to see from this self-consistency equation that $f_m = f_{m+N/2}$. Substituting (\ref{eqBAAT2}) into this result and performing some simplifications, we obtain the BA (Bethe ansatz) equation
\begin{equation}
    \frac{\beta_{r_{m-2k}} \pm \beta_{r_{m-2k}}^{-1}}{\beta_{r_{m-2k}} \mp \beta_{r_{m-2k}}^{-1}}
    - \frac{\beta_{r_{m}} \pm \beta_{r_{m}}^{-1}}{\beta_{r_{m}} \mp \beta_{r_{m}}^{-1}}
    =\frac{2}{\theta_{r_m}}. 
    \label{eqBAE}
\end{equation}
In the above simplification, we used the following fact: the direction of $\vec{\Delta x}$ corresponds to the symmetry axis $s_{-k}$, and thus we have $\beta_{s_{k-m}} = \beta_{s_{-k}s_{k-m}} = \beta_{r_{m - 2k}}$. 
When the BA equations (\ref{eqBAE}) hold, we can obtain $f_m = (\beta_{r_m} \mp \beta_{r_m}^{-1})/(\beta_{r_{m-2k}} \mp \beta_{r_{m-2k}}^{-1})$.
Substituting  this expression  into (\ref{eqConsistent}), one can observe  its validity.  
This demonstrates the equivalence between the BA equations (\ref{eqBAE}) and the self-consistency condition (\ref{eqConsistent}).

The Bethe ansatz equation (\ref{eqBAE}) is derived by using  $f_m = f_{m+N/2}$.
it appears that there are $N/2$ independent equations under this condition
. However, due to the definition of $\theta_g$, we can  verify $\sum_{n} \theta_{r_{m-2nk}}^{-1}=0$,  where the range of the summation is $n = 0, 1, \ldots, N/2 - 1$. 
The sum  all the right-hand sides of the  equations (\ref{eqBAE}) thus vanishes,  the sum on all  left-hand sides is trivially  zero. 
Hence, Eq. (\ref{eqBAE}) actually contains $N/2 - 1$ independent equations. 
Our derivation is general, so (\ref{eqBAE}) holds for any even $N$. 
The Bethe ansatz (\ref{eqBA}) involves  only two complex degrees of freedom. 
For $D_{2N}$ with $N\ge 4$, however,  the  Bethe ansatz equations
\ref{eqBAE} are  generically overconstrained:   the number of independent algebraic equations  exceeds that of  unknowns  $(k_0,\theta)$.
 Even if accidental solutions exist for finely tuned values of the parameter $c$, such solutions  form at most a measure-zero subset in the $(k_0,\theta)$–plane and do not yield the  robust, continuously parametrized families linked to integrability. 
 In this sense, for $D_{2N}$ ($N\ge 4$)  the model is  non-solvable  via   the coordinate Bethe ansatz.

\subsection{The quantum number and the solution of Bethe ansatz equations}

Let us  further  discuss the   solution  of the LQZC model with $N = 6$, $k = 1$. 
Now $\vec{\lambda}_\perp=[\cos\frac{2\pi}{3}, \sin\frac{2\pi}{3}]$, $r_m\vec{k}=k_0[\cos(\frac{\pi m}{3}+\theta), \sin(\frac{\pi m}{3}+\theta)]$ and the displacement $\vec{\Delta x} = \frac{L}{2}[\cos\frac{5\pi}{6}, \sin\frac{5\pi}{6}]$. After a lengthy calculation, we obtain the BA (Bethe ansatz) equations  (\ref{eqBAA})  in terms of the original $k_x$ and $k_y$
\begin{align}
    &\begin{cases}
        k_x + \sqrt{3}k_y = \frac{c}{2}\cot\frac{L}{4}(\sqrt{3}k_x+k_y)+\frac{c}{2}\cot\frac{L}{2}k_y,  \\
        -k_x + \sqrt{3}k_y = \frac{c}{2}\cot\frac{L}{4}(-\sqrt{3}k_x+k_y)+\frac{c}{2}\cot\frac{L}{2}k_y, 
    \end{cases}
    \quad \text{for even case; } \notag \\
    &\begin{cases}
        -k_x - \sqrt{3}k_y = \frac{c}{2}\tan\frac{L}{4}(\sqrt{3}k_x+k_y)+\frac{c}{2}\tan\frac{L}{2}k_y,  \\
        k_x - \sqrt{3}k_y = \frac{c}{2}\tan\frac{L}{4}(-\sqrt{3}k_x+k_y)+\frac{c}{2}\tan\frac{L}{2}k_y. 
    \end{cases}
    \quad \quad\text{for odd case.}
    \label{eqBAEL}
\end{align}
This coincides with with the BA equations obtained in \cite{liu2019}.  

However, obtaining the BA equations alone is far from sufficient to solve the problem. 
First, solving these equations is generally highly challenging. 
More critically, a set of quantum numbers must be found to exhaustively and uniquely enumerate the original model’s solutions-only then can we confidently assert the model is solved via the Bethe ansatz, as exemplified by the XXX spin model and Lieb-Liniger model \cite{bethe1931, lieb1963, lieb1963a}. 
Quantum numbers further enable analysis of the thermodynamics and dynamics of quantum integrable models, advancing BA method research to a deeper level.
From this standpoint, equations (\ref{eqBAE}) and (\ref{eqBAEL}) are insufficient to solve our model. 
While these equations admit infinitely many solutions, we lack a systematic way to enumerate all model solutions, solve the equations themselves, or even identify the model’s ground state.

\begin{figure}[t]       
  \centering               %
  \includegraphics[width=1.0\textwidth]{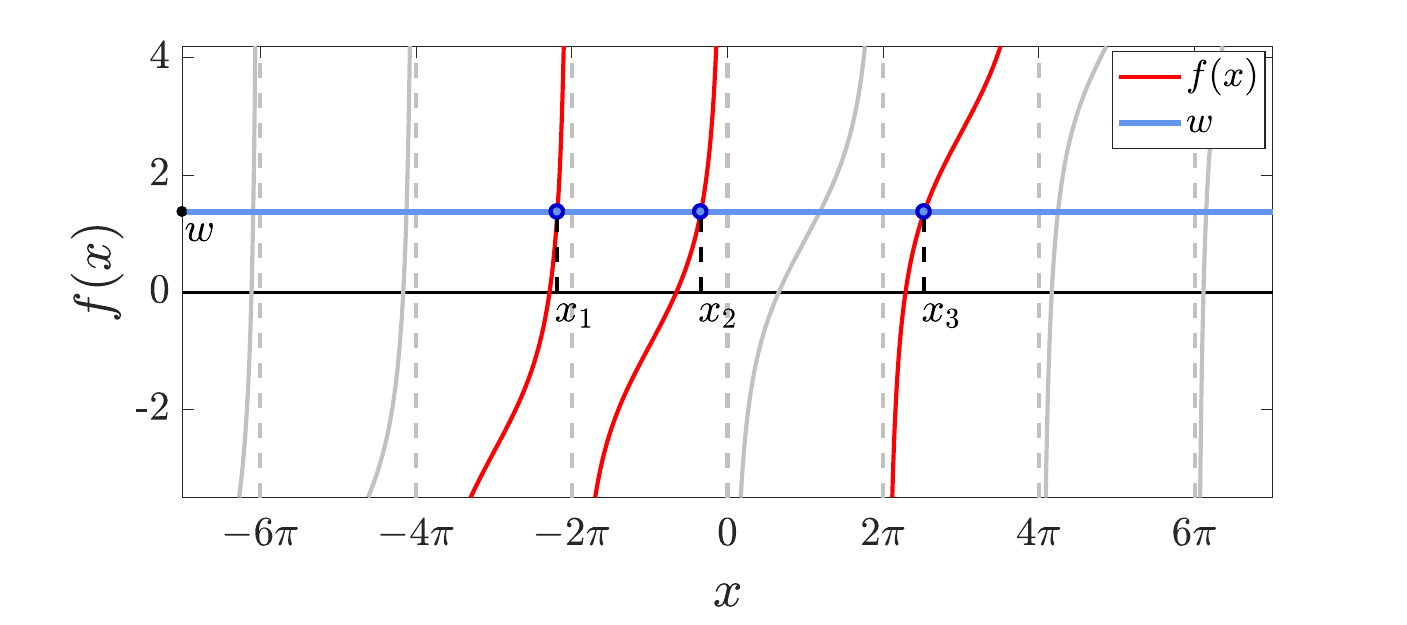} 
  \caption{Example of numerical solution of the equation (\ref{eqBAESep}): for a fixed set of $(n_1, n_2, n_3) = (-2, -1, 1)$, we only need to search for $w=f(x_i)$ such that $x_1 + x_2 + x_3 = 0$ holds, thereby obtaining the unique solution of the BA equations  associated with this set of integers.}
  \label{figffunction}       %
\end{figure}

In order to identify the quantum numbers and systematically solve for all solutions of this model, we perform a transformation on the BA equation (\ref{eqBAE}). We introduce the following reparameterization:
\begin{equation}
    x_1 = k_0L\cos(\theta-\frac{\pi}{6}), \quad 
    x_2 = k_0L\cos(\theta+\frac{\pi}{2}), \quad
    x_3 = k_0L\cos(\theta + \frac{7\pi}{6}),
    \label{eqx123}
\end{equation}
define $\eta=\frac{2}{cL\sin(2\pi/3)}$, we obtain the ``separated variables'' type of BA equations:
\begin{align}
    \eta x_1 \mp \cot^\pm\frac{x_1}{2}
    &= \eta x_2 \mp \cot^\pm\frac{x_2}{2}
    = \eta x_3 \mp \cot^\pm\frac{x_3}{2}, \notag \\
    &  x_1 + x_2 + x_3 = 0. 
    \label{eqBAESep}
\end{align}
To obtain the quantum numbers, we rewrite the BA equations for the even parity as  the following form  (similarly for the odd parity case)
\begin{equation}
    f(x)=\eta x - \cot \frac{x}{2}.
    \label{eqf}
\end{equation}
Here  we  assume $c > 0$. 
All integer multiples of $2\pi$ 
 are singularities of this function. We thus restrict our search to  $x_i \in (2\pi n_i, 2\pi (n_i+1))$ for $i=1,2,3$ ( where $n_i$ are integers). 
With$n_i$  fixed, let $w=f(x_i) $, we only need to perform a bisection search along the $y$-axis to find $w$ under the condition $x_1 + x_2 + x_3 = 0$. 
This yields  the solution to the  equation (\ref{eqf}) for the integer set  $(n_1, n_2, n_3)$. 
Figure \ref{figffunction} shows  the  numerical solution for these  fixed integers. 
From the obtained  $x_{1,2,3}$, we find the solution of BA as 
$k_0 = \sqrt{\frac{2(x_1^2+x_2^2+x_3^2)}{3L^2}}$ and $\theta = \frac{\pi}{6} + \arccos\frac{x_1}{k_0L}$, which recover the solution given in \cite{liu2019}.

In fact, the set of integers introduced above is not the set of quantum numbers which we are searching for.
 The later,  must satisfy the constraint $-3 < n_1 + n_2 + n_3 < 0$ in order to admit a solution  for the equation (\ref{eqBAESep}).
 By analyzing the solutions, we observe that for  given two positive integers $p$ and $q$ with $p < q$  in  the even-parity case, we have
\begin{equation}
    n_1=\left\lfloor \frac{p-q}{3}\right \rfloor - p, \quad  
    n_2=\left\lfloor \frac{p-q}{3}\right \rfloor, \quad 
    n_3=\left\lfloor \frac{p-q}{3}\right \rfloor + q, \quad  
    \text{s.t. } p - q \not\equiv 0 \pmod{3};
    \label{eqQuantumNumberEven}
\end{equation}
and for the odd-parity case, we have:
\begin{equation}
    n_1=\left\lfloor \frac{p-q}{3} + \frac{1}{2}\right \rfloor - p, \quad  
    n_2=\left\lfloor \frac{p-q}{3} + \frac{1}{2}\right \rfloor, \quad 
    n_3=\left\lfloor \frac{p-q}{3} + \frac{1}{2}\right \rfloor + q .
    \label{eqQuantumNumberOdd}
\end{equation}
Here the notation $\lfloor \, \rfloor$ denotes the floor function. 
The integers $p$ and $q$ are the quantum numbers which we seek.  
For  $1 \leq  p < q$, we use (\ref{eqQuantumNumberEven}) or (\ref{eqQuantumNumberOdd}) to find  $n_{1,2,3}$ under the  constraint  $x_i \in (2\pi n_i, 2\pi (n_i+1))$,
 yielding the unique solution to the separated-variable BA equation (\ref{eqBAESep}).
This quantum number choice is consistent with that for hard-core bosons, i.e.  model reduces to hard-core bosons as $c \to \infty$.
Moreover,  the additional constraint $p - q \not\equiv 0 \pmod{3}$ on quantum numbers in this even-parity case implies our Bethe ansatz equations may miss some solutions. 
We address this issue in the subsequent numerical study.

\subsection{The numerical solutions}

In this section, we employ the finite element method (FEM) to solve the model.
The FEM is a numerical technique that partitions a complex continuum into simple finite elements, constructs approximate equations on each element, and assembles them into a global system of equations \cite{pepper2005, brenner2008, hughes2012}.
This approach offers an efficient means of solving partial differential equations and related problems in engineering and physics.
For the present model, an additional aspect requiring special treatment is the $\delta$-function barrier.
Appendix A provides a detailed introduction to the FEM tailored to the specific problem under consideration.

\begin{figure}[t]       
  \centering               %
  \includegraphics[width=0.85\textwidth]{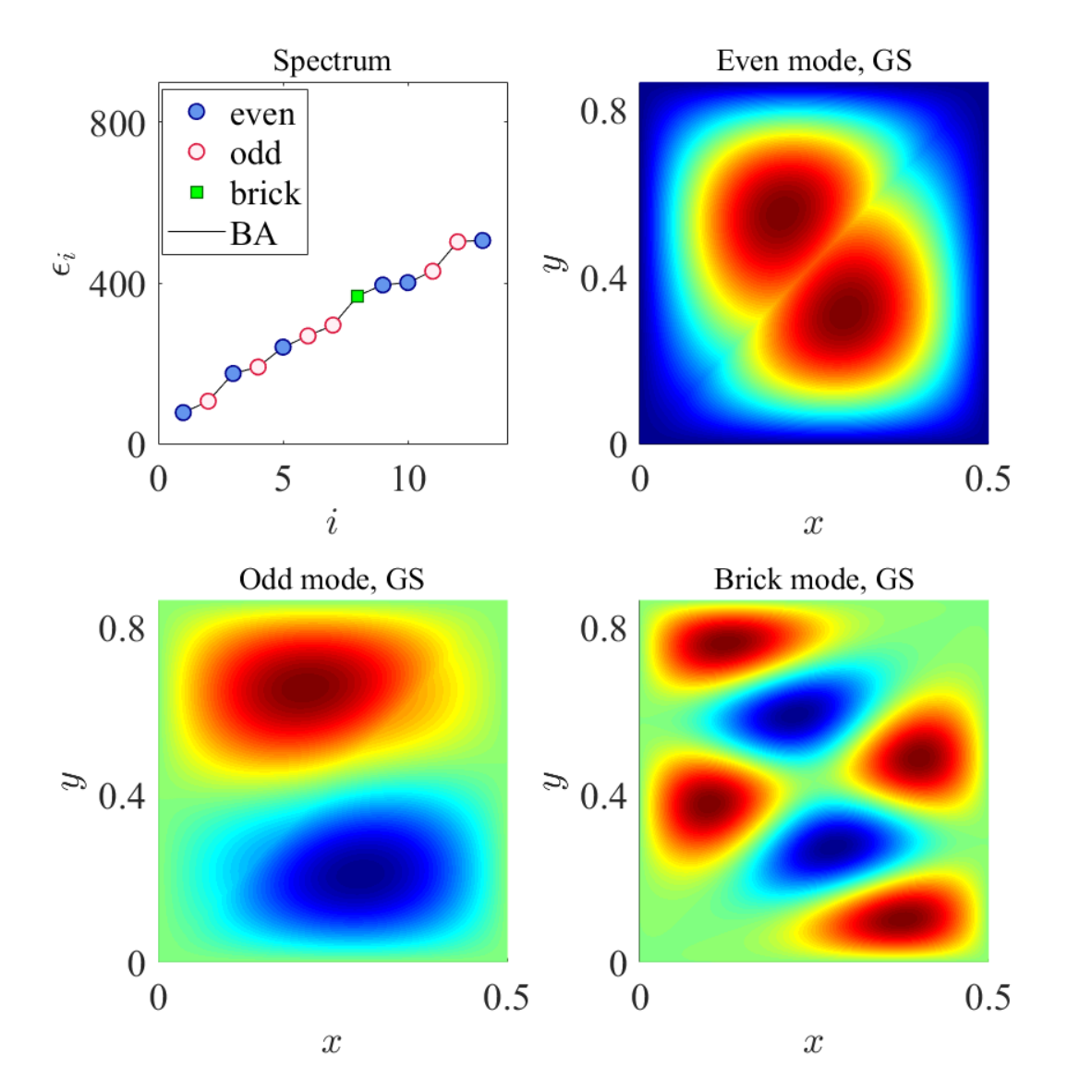} 
  \caption{The numerical solutions of the LQZC model are obtained using the FEM method. We set the parameters $L = 1$ and $c = 10.0$. The three contour plots in the figure show the ground-state solutions of the three different modes.}
  \label{figSolutionLQZC}       %
\end{figure}

In Figure \ref{figSolutionLQZC}, we show our numerical  results.
 The upper left panel shows the  numerical  energy spectrum which  agrees very well with the spectrum obtained from the Bethe ansatz solution. 
 The horizontal axis  $i$ denotes the energy level index.
  In the remaining three panels, the $z$-coordinate of the contour plots represents the wave function value. 
  We note that, in addition to even- (upper right panel) and odd-parity (lower left panel) solutions, we find a class of solutions not captured by the Bethe ansatz equations (\ref{eqBAESep}), denoted by green points in the upper left panel’s energy spectrum—termed the brick mode". 
  Such solutions were previously reported in \cite{jackson2024}. Notably, the mode is constructed by introducing $\delta$-function barriers into the original domain, partitioning it into smaller right triangles (lower left panel). It is formed by assembling solutions to these right triangles via reflection symmetry--hence the name brick mode".
  
  At the end of the preceding subsection, we noted that even-parity wave functions impose an additional constraint on the quantum numbers $p$ and $q$, implying some solutions are missing from the Bethe ansatz spectrum. 
  We emphasize these missing solutions are precisely the "brick mode" states.
We would like to  point out that these missing solutions are precisely the "brick mode" states. 
These modes can still be characterized by the quantum numbers $p$ and $q$ for $p<q$
\begin{align}
     n_1&=-\frac{2p+q}{3}, \quad  n_2=\frac{p-q}{3}, \quad n_3=\frac{p+2q}{3}, 
     \quad \text{s.t. }  p - q \equiv 0 \pmod{3}; \notag \\
     &E_{\text{brick}} = \frac{8\pi^2}{3L^2}\left(  n_1^2+n_2^2+n_3^2 \right).
    \label{eqQuantumNumberBrick}
\end{align}
It is clear that the energy of this mode no longer depends on $c$, since the value of the wave function is strictly zero on the barrier.

For quantum numbers $1 \leq p < q$,  equations (\ref{eqQuantumNumberEven}) and (\ref{eqQuantumNumberBrick}) yield two solutions: one for odd parity, and the other corresponding to either even parity or the brick mode.
In summary, strong evidence indicates the BA equations (\ref{eqBAESep}) lack completeness--supplemented by the brick mode.
 We recover the full energy spectrum of the LQZC model. 
 This completeness is further corroborated by our numerical results.
Here, the brick mode acts as a singular solution to BA Equation (\ref{eqBAESep}), i.e., when both sides of the equation diverge to 
$\pm \infty$. 
Nonetheless, it must be recovered via methods beyond direct solution of the BA equation.
The completeness of BA equations is inherently a subtle issue. 
They may not capture all possible solutions, and systematically recovering the missing ones via simple methods is often challenging.
 In subsequent sections, we present more specific examples to illustrate this point.

\section{Gaudin's Kaleidoscope model and the Kaleidoscope Yang-Baxter Equation}

In Ref. \cite{jackson2024}, the authors astutely note that the LQZC model’s integrability is nontrivial, as it violates Gaudin’s necessary conditions for integrability. 
To resolve this, the authors creatively extended these conditions to accommodate the LQZC model within a new framework. 
However, this approach raises conceptual difficulties.
Firstly, the LQZC model itself possesses no reflection symmetry other than the $\mathbb{Z}_2$ arising from the $\pi$-rotation symmetry discussed above. 
Moreover, as highlighted in the previous section, brick mode solutions rely on introducing additional barriers to partition the model.
 For general solutions, we cannot conclude whether this argument remains valid.

\subsection{Gaudin's approach revisited}
To fully understand the ideas in \cite{jackson2024}, we closely revisited Gaudin's book. We found that Gaudin's original discussion includes specific constraints. 
Here we quote Gaudin's statement (p. 84) concerning Eq. (5.32) \cite{gaudin2014}.
\begin{quote}
A restrictive hypothesis is made, namely that the function $\psi(x)$ is \textit{symmetric} in $\mathbb{R}_N$, in other words that its data in $D$ suffices to determine it in the whole of $\mathbb{R}_N$ using the properties (5.28):
    \begin{equation}
        \psi_{\{k\}}(gx)=\psi_{\{k \}}(x), \quad x\in D.  \tag{5.32}
    \end{equation}
\end{quote}
This is, of course, a very strong restriction, as it limits the wave function’s symmetry group to a nontrivial representation.
Gaudin’s motivation for this choice was to relate the Bethe ansatz solution to the root systems of Lie algebras, i.e.  a classical problem in Euclidean geometry.
Gaudin was clearly aware of the specialized nature of this discussion. 
At the end of Section 5.2 (p. 89), he noted 
\begin{quote}
    We shall moreover not mention anything of a more interesting problem consisting of constructing the wave-functions belonging to an irreducible representation of $G$ of dimension higher than $1$:
    \begin{equation*}
        \psi(gx)=T(g)\psi(x), 
    \end{equation*}
    in other words having an arbitrary `type of symmetry’, except for algebra $A_{N-1}$ corresponding to the permutation group $\pi_N$ whose problem is treated completely in Chapters 11 and 12.
\end{quote}

Clearly, this approach raises a key problem: the integrability conditions from Gaudin’s method only determine if the model is Bethe-ansatz solvable within a specific symmetry sector.
Gaudin’s integrability is thus best understood as a sufficient condition for Bethe-ansatz solvable states in a representation sector, not for the full many-body spectrum.
We emphasize that \textbf{a model may be Bethe-ansatz solvable only in specific symmetry subspaces, but not generically.}

\begin{figure}[t]       
  \centering               %
  \includegraphics[width=0.8\textwidth]{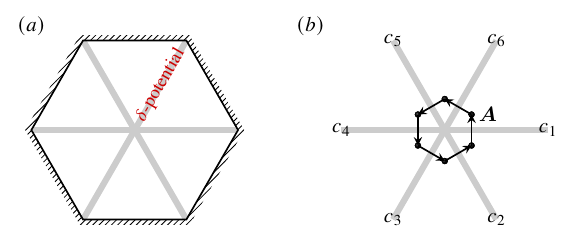} 
  \caption{(a) Gaudin's kaleidoscope model with $D_6$ symmetry. Within the regular hexagonal hard-wall potential, three $\delta$-function barriers are placed along the three diagonals of the hexagon, each with the same strength.}
  \label{figGKModel}       %
\end{figure}

\subsection{Gaudin's $D_6$ kaleidoscope model: symmetry-resolved analysis}

To elaborate further, we next consider a fundamental example: Gaudin’s kaleidoscope model with  $D_6$  symmetry.
To ensure the eigenvalue problem is well-defined, we impose hard-wall boundary conditions on the regular hexagon in Fig. \ref{figGKModel}(a).
This model exhibits perfect  $D_6$ symmetry: the Hamiltonian, boundary conditions, and Bethe ansatz method all satisfy its symmetry requirements.
We incorporate symmetry into the FEM approach by solving the Hamiltonian in each symmetry subspace.To this end, we first list the conjugacy classes of  $D_6$, whose character table is given by

\vspace*{0.2cm}
\hspace*{0.5cm}
\noindent
\begin{minipage}[c][0.2\textheight][c]{0.58\textwidth} 
  \hspace*{1.5cm}
  \vspace*{-0.7cm}
  \begin{tabular}{|c|r|r|r|r|r|r|}
    \hline
    $D_6$ & $\mathcal{C}_1$ & $\mathcal{C}_2$ & $\mathcal{C}_3$ & $\mathcal{C}_4$ & $\mathcal{C}_5$ & $\mathcal{C}_6$ \\  
    \hline
    $\Gamma_1$ & 1  & 1 & 1& 1& 1& 1\\
    $\Gamma_2$ & 1  & 1 & 1& 1& -1& -1\\
    $\Gamma_3$ & 1  & -1 & -1& 1& 1& -1\\
    $\Gamma_4$ & 1  & -1 & -1& 1& -1& 1\\
    $\Gamma_5$ & 2  & -2 & 1& -1& 0& 0\\
    $\Gamma_6$ & 2  & 2 & -1& -1& 0& 0\\
    \hline
\end{tabular}
 \\[28pt] 
  \hspace*{1cm} \textbf{  Table 1: Character table of the $D_6$ group}
\end{minipage}%
\hspace*{-1.5cm}
\begin{minipage}[c][0.2\textheight][c]{0.38\textwidth}
  \vspace*{-0.6cm}
 \begin{align*}
\mathcal{C}_1 &= \{r_0 \} \\
\mathcal{C}_2 &= \{r_3 \} \\
\mathcal{C}_3 &= \{r_1,r_5 \} \\
\mathcal{C}_4 &= \{r_2,r_4 \} \\
\mathcal{C}_5 &= \{s_0, s_2, s_4 \} \\
\mathcal{C}_6 &= \{s_1, s_3, s_5 \} \\
\end{align*}
\end{minipage}

According to  group representation theory, the original space of wave functions can be projected out using projection operators, which are given by the character table as follows:
\begin{equation}
    P^{\alpha} = \frac{d_\alpha}{\vert D_6 \vert}\sum_{g \in D_6} \chi^{(\alpha)}(g)U(g).
    \label{eqProjection}
\end{equation}
Here, $\alpha$ labels the conjugacy classes, corresponding one-to-one with the irreducible representations; $d_\alpha$ is the dimension of the representation; $U(g)$ is the operator acting on the symmetric space, which, in this context, corresponds to the rotation and reflection operations on the wave function.

By utilizing the irreducible representations of this symmetry group, we can decompose the wave function according to its symmetry properties. 
For a specified symmetry $\Gamma$, the wave function can be rewritten as:
\begin{equation}
    \Psi_\Gamma(\vec{x})=\sum_a M^{\Gamma}_{a,\eta} \Psi_a(\eta^{-1} \vec{x}), \text{  if } \eta^{-1}\vec{x} \text{ in } D_F.
    \label{eqDecomposition}
\end{equation}
Here, $D_F$ denotes the fundamental region, which, specifically, is a right triangle with an acute angle of $\pi/6$. 
By applying the operations of the $D_6$ group on this right triangle, one can generate the entire regular hexagon. $a = 1$ or $a = \{1,2\}$ corresponds to the one- or two-dimensional irreducible representations, respectively. 
By the projection operator (\ref{eqDecomposition}), we can obtain the explicit form of the matrix $M^\Gamma_{a, \eta}$ as follows (treat the group elements as column indices):
\begin{align}
    M^{\Gamma_1} &= [1,1,1,1,1,1,1,1,1,1,1,1], \notag \\
    M^{\Gamma_2} &= [1,1,1,1,1,1,-1,-1,-1,-1,-1,-1], \notag \\
    M^{\Gamma_3} &=[1,-1,1,-1,1,-1,1,-1,1,-1,1,-1], \notag \\
    M^{\Gamma_4} &=[1,-1,1,-1,1,-1,-1,1,-1,1,-1,1], \notag \\
    M^{\Gamma_5^{(1)}}_1 & = [0,-1,-1,0,1,1,0,-1,-1,0,1,1]/\sqrt{8}, \notag \\
    M^{\Gamma_5^{(1)}}_2 &= [2,1,-1,-2,-1,1,2,1,-1,-2,-1,1]/\sqrt{24},  \notag \\
     M^{\Gamma_5^{(2)}}_1 & = [2,1,-1,-2,-1,1,-2,-1,1,2,1,-1]/\sqrt{24}, \notag  \\
    M^{\Gamma_5^{(2)}}_2 &= [0,1,1,0,-1,-1,0,-1,-1,0,1,1]/\sqrt{8},  \notag \\
     M^{\Gamma_6^{(1)}}_1 & = [0,-1,1,0,-1,1,0,-1,1,0,-1,1]/\sqrt{8}, \notag \\
    M^{\Gamma_6^{(1)}}_2 &= [2,-1,-1,2,-1,-1,2,-1,-1,2,-1,-1]/\sqrt{24},  \notag \\
     M^{\Gamma_6^{(2)}}_1 & = [2,-1,-1,2,-1,-1,-2,1,1,-2,1,1]/\sqrt{24}, \notag \\
    M^{\Gamma_6^{(2)}}_2 &=[0,1,-1,0,1,-1,0,-1,1,0,-1,1]/\sqrt{8}.  
    \label{eqGammaValue}
\end{align}
$\Gamma_{1,2,3,4}$ correspond to the four one-dimensional irreducible representations of the $D_6$ group.
For its two 2D representations, each has two distinct $\Gamma_{5, 6}$ choices (i.e., each 2D representation has multiplicity two).
Accounting for symmetry, our FEM calculations are restricted to the fundamental region-only $1/12$ of the original domain.
Given that the computational cost of eigenvalue problems typically scales as $O(N^3)$, this symmetry-based decomposition yields a ~$10^3$-fold computational speedup.
The only caveat is that at the boundaries of the fundamental region, the symmetry group reduces from $D_6$ to ${Z}_6$ requiring careful handling.
Since this work does not focus on numerical computation, we do not elaborate on these technical details in detail.

\begin{figure}[t]       
  \centering               %
  \includegraphics[width=0.98\textwidth]{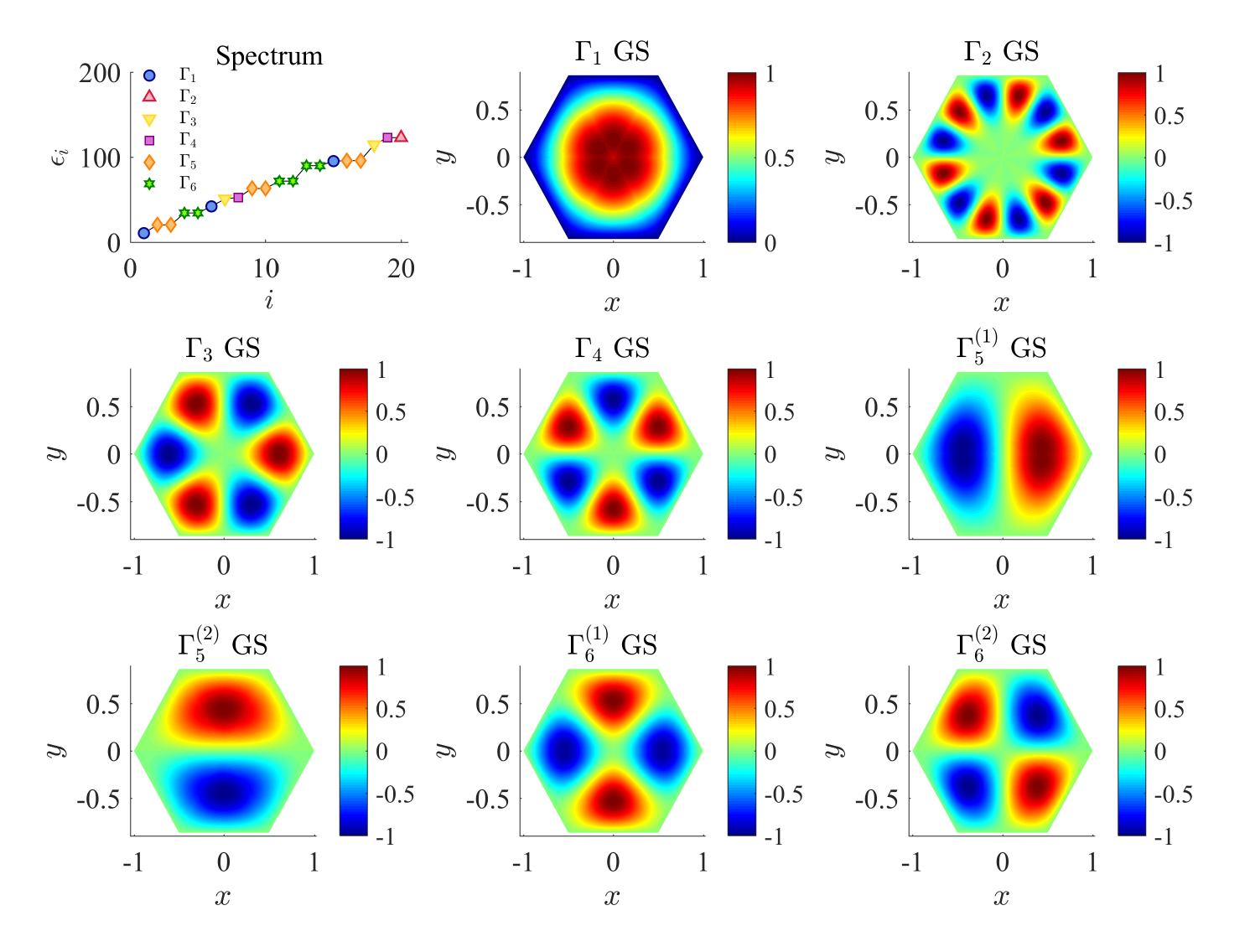} 
  \caption{The Kaleidoscope Gaudin model solved by FEM. The top-left figure shows the energy spectrum, while the remaining figures display the ground-state solutions in each symmetry subspace. Here we take $c = 1.0$, and the entire computational domain is partitioned into $89,448$ small triangles (elements). All wave functions are normalized so that their maximum value is $1$, for the purpose  of a good  visualization.}
  \label{figKGSolution}       %
\end{figure}

We present our FEM results in Fig. \ref{figKGSolution}.
The ground state resides in the subspace of the trivial representation.I
n contrast, the antisymmetric representation exhibits the highest energy of the ground state.
While not strictly proven, we observe that wave functions with more nodes typically correspond to higher energies.

While numerical solutions offer many advantages, the main focus of this work is on the discussion of integrability.
In what follows, we present a numerical algorithm to check if this model is solvable for a specific symmetry using the coordinate Bethe ansatz.
The Bethe ansatz reads
\begin{equation}
    \Psi(\vec{x}) =\sum_g A_g\mathrm{e}^{\mathrm{i}(\vec{x},g\vec{k})},
    \quad \vec{k}=[k_0 \cos\theta, k_0 \sin\theta].
    \label{eqTry}
\end{equation}
as a type of trial variational wave function.
Here the variational parameters include $A_g$, $k_0$ and $\theta$. 
Let us  denote the numerical wave function $\Psi_0$ that has already been obtained (it has been numerically solved and properly normalized).
If $\Psi_0$ is representable by the Bethe ansatz form, then the maximal normalized overlap should approach $1$ upon optimizing the variational parameters namely, 
\begin{equation}
    \max_{\scriptscriptstyle \langle \Psi_\mathrm{BA}\vert\Psi_\mathrm{BA}\rangle=1}
    \vert \langle \Psi_0\vert\Psi_\mathrm{BA}\rangle \vert \to 1.
    \label{eqOverlapLimit}
\end{equation}
It is clear that $k_0 = \sqrt{E}$ must be satisfied. 
Therefore we first chose to fix $k_0$, and then adjust the other parameters to examine $(\ref{eqOverlapLimit}$)  whether the above relation can be fulfilled.

It is not difficult to show that, once $\theta$ is fixed, the remaining optimization problem with respect to the left hand side of  (\ref{eqOverlapLimit}) reduces to a quadratic optimization over the parameters $A_g$. The corresponding loss function is defined as follows:
\begin{equation}
    \mathcal{L}[\bm{A}] = \bm{B}^\dagger\bm{A} - \beta(\bm{A}^\dagger\bm{M}\bm{A}-1)
    \label{eqLagrangian}
\end{equation}
with $\beta$ being the Lagrange multiplier, and matrices $\bm{M}$ and column vector $\bm{B}$ are defined by the integrals over the fundamental region as the following:
\begin{equation}
    M_{g^\prime g}=\int_{D_F}\mathrm{d}\vec{x} \;\mathrm{e}^{\mathrm{i}(g\vec{k}-g^\prime \vec{k}, \vec{x})}, \quad B_{g}=\int_{D_F}\mathrm{d}\vec{x} \;\Psi_0(\vec{x})\mathrm{e}^{\mathrm{i}(g\vec{k}, \vec{x})}.
    \label{eqMandB}
\end{equation}
Thus, the maximal overlap between the two wave functions and its deviation from 1 can be obtained as:
\begin{equation}
    1 - \max_{\scriptscriptstyle \langle \Psi_\mathrm{BA}\vert\Psi_\mathrm{BA}\rangle=1} 
    \vert \langle \Psi_0\vert\Psi_\mathrm{BA}\rangle \vert = 1 - \bm{B}^\dagger \bm{M}^{-1}\bm{B}. 
    \label{eqMinDis}
\end{equation}

\begin{figure}[t]       
  \centering               %
  \includegraphics[width=0.92\textwidth]{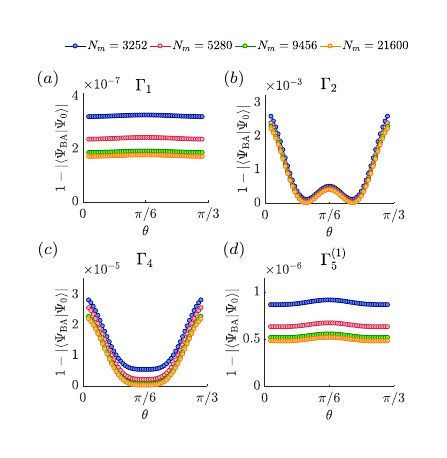} 
  \caption{For numerically obtained ground-state wave functions in different symmetry sectors, we use the Bethe ansatz as a trial wave function and plot the results of Equation (\ref{eqMinDis}) as a function of $\theta$.
   The different colored lines represent  the results  for different numbers of elements which we used in the FEM algorithm, see the legend. }
  \label{figKGFit}       %
\end{figure}

By running over $\theta$, we calculate  the corresponding $\bm{M}$ and $\bm{B}$ matrices, and then compute  the minimal error given by Equation (\ref{eqMinDis}), we can examine how this error depends on $\theta$. This dependence serves as a criterion for whether the original wave function can be represented by the Bethe ansatz (\ref{eqTry}) regarded as a trial wave function.

\begin{wrapfigure}{r}{0.25\textwidth} 
    \centering
    \includegraphics[width=0.25\textwidth]{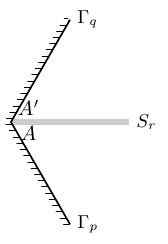} 
     \caption{Depict the symmetry with respect to the Eq. (\ref{eqBAALocal}).
     }
    \label{Addtional}
\end{wrapfigure}

 In Figure \ref{figKGFit}, we show the dependence of the minimal distance on $\theta$. The sub-figures correspond to the four representative ground states, $\Gamma_1$, $\Gamma_2$, $\Gamma_4$, and $\Gamma_5^{(1)}$, as indicated in Figure \ref{figKGSolution}.
We observe that in panels (b) and (c) in Fig.~ \ref{figKGFit}, the minimal distance is sensitive to $\theta$ and approaches zero at specific values of $\theta$.
 In contrast, for panels (a) and (d), the minimal distance shows no significant dependence on $\theta$, and there is no indication that it approaches zero as the number of elements increases.
  In fact, according to our numerical results, among all eight symmetry sectors, only $\Gamma_2$ and $\Gamma_4$, corresponding to Figure \ref{figKGFit}(b) and \ref{figKGFit}(c),  are integrable via the Bethe ansatz.
   The remaining six sectors, including $\Gamma_1$ who contains the ground state of the system, cannot be solved by the Bethe ansatz.

The calculations above confirm the opening statement: the integrability of the coordinate Bethe ansatz depends on the decomposition into irreducible representations of the symmetry group. 
Even if we find solutions using the coordinate Bethe ansatz, it does not guarantee that we can obtain all solutions, or even key solutions like the ground state. 
In the example above, the solutions obtained via the coordinate Bethe ansatz are only a small subset of all possible solutions.

We can analytically prove the observations from our numerical calculations. In the present model, the barrier forms an angle of  $\pi/3$  with the hard wall. 
According to Ref. \cite{jackson2024}, its integrability is guaranteed by the asymmetric Bethe ansatz. 
We further conclude that Bethe ansatz solutions exist only in the two symmetry sectors $\Gamma_2$ and $\Gamma_4$.
 Following the local structure of the model as depicted and the rules established in Section \ref{sec2}, we readily obtain the following equations:

\begin{equation}
        \begin{cases}
            \bm{A}^\prime=\bm{S}_r\bm{A} ,\\
            (\bm{I} + \bm{\Gamma}_p)\bm{A}=0 ,\\
            (\bm{I} + \bm{\Gamma}_q)\bm{A}^\prime=0 .
        \end{cases}
        \label{eqBAALocal}
\end{equation}
Let $\bm{A} = [\bm{X}; \bm{Y}]$ ($\bm{X}$ corresponds to the rotational group elements, while $\bm{Y}$ corresponds to the reflection group elements), and by simplifying the above equations, without loss of generality setting $r = 0$, we obtain the following linear equation:

\begin{equation}
    \left(\bm{I} - \bm{t}^{p-q} + \frac{c}{2\mathrm{i}k_0}(\bm{I}-\bm{t}^{-q})\bm{s}(1-\bm{t}^p)   \right)\bm{X} = 0,  
    \label{eqLinear}
\end{equation}
with $\bm{t}$ and $\bm{s}$ defined in (\ref{eqts}).
Here we  consider  $p = -2$ and $q = 2$. 
Substituting these values into Equation (\ref{eqLinear}) with lengthy calculation, we find the nontrivial solutions of this equation indicating  the following condition
\begin{equation}
    \bm{X} = \bm{t}^2\bm{X},
\end{equation}
This shows  that the solutions of Equation (\ref{eqBAALocal}) are antisymmetric with respect to the symmetry axis $s_r$. 
In other words, for the local structure of the model depicted in the figure Fig.~\ref{Addtional}, any solution obtained via the Bethe ansatz must vanish exactly on the $\delta$ barrier. These solutions correspond precisely to the two symmetry sectors $\Gamma_2$ and $\Gamma_4$.

\subsection{The Kaleidoscope Yang-Baxter Equation}
In Section 2, we established the rules for constructing sections, and in Section 3, we applied these rules to solve the LQZC model. 
However, for Gaudin’s Kaleidoscope model considered in this section,  an additional self-consistency condition is required. 
As being  illustrated in Figure \ref{figGKModel}(b), for a section $A$ near the origin, after undergoing a sequence of scatterings, it must return to itself. 
This can be expressed by  $\bm{S}_0\bm{S}_{10}\bm{S}_8\bm{S}_6\bm{S}_4\bm{S}_2\bm{A} = \bm{A}$. 
We provide a sufficient condition for the validity of the above equation for the case of even $N$:
\begin{equation}
    [\bm{T}\bm{S}_0(\theta,z)]^N = \bm{I}
    \label{eqKYBE}
\end{equation}
with matrices $\bm{T}$ and $\bm{S}_0$ defined in (\ref{eqTS}), we denote  $z = \frac{c}{2\mathrm{i}k_0}$ and treat  $\bm{S}_0$ as a function of $\theta$ and $z$. 
Like the Yang–Baxter Equation (YBE), (\ref{eqKYBE}) ensures the scattering matrix’s self-consistency (a sufficient condition). 
By analogy, we call this the Kaleidoscope Yang–Baxter Equation (KYBE) as it plays a role analogous to YBE in enforcing multi-scattering consistency, although its algebraic form differs from the standard braid-type YBE. 
Similar to the YBE, it involves two complex parameters.

Equation (\ref{eqKYBE}) is nontrivial, primarily due to $\bm{S}_0$ nonlinear dependence on $\theta$.
 Admittedly, we lack a sufficiently concise proof for it (unlike standard proofs for YBE with given $R$-matrices). 
 Below, we outline a more involved proof. We prove the following sufficient condition: the eigenvalues of 
 $\bm{T}\bm{S}_0$ are $\exp(2\pi\mathrm{i} n / N)$ for $n = 0, 1, \ldots, N-1$
  each with multiplicity two. 
  To this end, we consider the eigenvalue problem
  $\bm{T}\bm{S}_0 \bm{A} = \omega^k \bm{A}$
   with  $\omega = \exp(2\pi\mathrm{i} / N)$.
   For nonzero solutions, we write  $\bm{A}$ in block matrix form, 
   $A = \begin{bmatrix} \bm{X} ;\bm{ Y} \end{bmatrix}$,  leading to the following equation for $\bm{X}$:

\begin{equation}
    (\bm{I} - \omega^k \bm{t})(\alpha_1 \bm{d} + \alpha_{-1}\bm{d}^{-1})(\bm{I} - \omega^k \bm{t}^{-1})\bm{X} = \omega^k (\bm{t} - \bm{t}^{-1})\bm{X}. 
    \label{eqEQ0}
\end{equation}
In the aove equation, we have reparameterized variable  $z$ and $\theta$ by $\alpha_1:=-\mathrm{i}z\mathrm{e}^{\mathrm{i}\theta}$ and $\alpha_{-1}=\mathrm{i}z\mathrm{e}^{-\mathrm{i}\theta}$.
We list the definition of the $\bm{t}$ and $\bm{d}$, as they play a crucial role in our subsequent analysis:
    \begin{equation}
        \bm{t} :={
        \begin{bmatrix}
            0 & 1 & 0 & \cdots & 0 \\
            0 & 0 & 1 & \cdots & 0 \\
            0 & 0 & 0 & \cdots & 0 \\
            \vdots & \vdots & \vdots & \ddots & \vdots \\
            1 & 0 & 0 & \cdots & 0
        \end{bmatrix}_{N \times N}}, \quad
        \bm{d} :={
        \begin{bmatrix}
            \mathrm{e}^{\frac{2\pi \mathrm{i}}{N}\cdot 0} & 0  & \cdots & 0 \\
            0 & \mathrm{e}^{\frac{2\pi \mathrm{i}}{N}\cdot 1}  & \cdots & 0 \\
            \vdots & \vdots & \ddots & 0 \\
            0 & 0 & \cdots & \mathrm{e}^{\frac{2\pi \mathrm{i}}{N}\cdot (N-1)}
        \end{bmatrix}_{N \times N}}.
        \label{eqtd}
    \end{equation}
Then applying the Fourier transformation with $U_{mn} = N^{-1/2}\exp(2\pi\mathrm{i}\cdot mn/N)$ to Equation (\ref{eqEQ0}), and using the relations $\bm{U} \bm{t} \bm{U}^\dagger = \bm{d}^{-1}$ and $\bm{U} \bm{d} \bm{U}^\dagger = \bm{t}$, we  obtain the Fourier transformation form
\begin{equation}
     \hat{H}\tilde{\bm{X}}=\left((\bm{I} - \omega^k \bm{d}^{-1})(\alpha_1 \bm{t} + \alpha_{-1}\bm{t}^{-1})(\bm{I} - \omega^k \bm{d})+\omega^k (\bm{d} - \bm{d}^{-1})\right)\bm{\tilde{X}} = 0.
    \label{eqEQ1}
\end{equation}
with  the Fourier transformation of $\bm{X}$ by $\tilde{\bm{X}} = \bm{U} \bm{X}$.
We note that the first term on the left-hand side of $\hat{H}$ in Equation (\ref{eqEQ1}) is a product of three simple matrices: the leftmost matrix is diagonal with the $(k+1)$-th row set to zero; the rightmost matrix is also diagonal with the $(N-k+1)$-th column set to zero and the middle matrix is a tridiagonal matrix. 
The second term is a diagonal matrix, in which the elements in the first row and the $(N/2 + 1)$-th row are zero.
\begin{figure}[t]       
  \centering               %
  \includegraphics[width=0.56\textwidth]{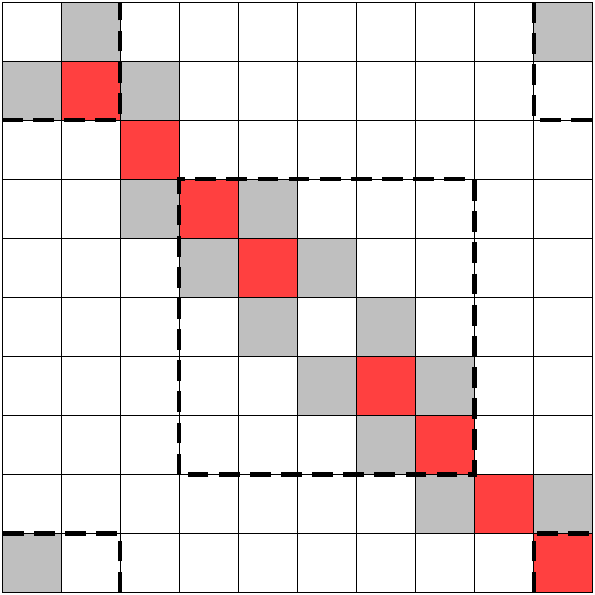} 
  \caption{A chessboard representation of the nonzero elements of the matrix $\hat{H}$ in Equation (\ref{eqEQ1}). }
  \label{figchessboard}       %
\end{figure}
We represent the distribution of the elements of $\hat{H}$ using a checkerboard diagram in Figure \ref{figchessboard}: white squares indicate zero elements, gray squares correspond to nonzero elements from the first term of $\hat{H}$, and red squares represent nonzero elements from the second term of $\hat{H}$. 
In this example, we take $N = 10$ and $k = 2$.
Note that in this example, the third row of the matrix is zero, which implies $X_3 = 0$. 
For the remaining components of $\bm{X}$, we divide them into two linear systems, as indicated by the two black dashed boxes in Figure \ref{figchessboard}.

We focus on the linear system highlighted by the black dashed box in the middle of the Figure \ref{figchessboard}. Its matrix can be obtained directly by projecting the matrix in Equation (\ref{eqEQ1}) onto the relevant subspace:
\begin{align}
     &\left[(\alpha_1 \bm{t} + \alpha_{-1}\bm{t}^{-1})_B+\omega^k 
     (\bm{I} - \omega^k \bm{d}^{-1})_B^{-1}(\bm{d} - \bm{d}^{-1})_B (\bm{I} - \omega^k  \bm{d})_B^{-1}\right]
     (\bm{I} - \omega^k  \bm{d})_B\bm{\tilde{X}}_B  \notag \\
      = & \,\, \hat{H}_B(\bm{I} - \omega^k  \bm{d})_B\bm{\tilde{X}}_B = 0.
    \label{eqEQ1-1}
\end{align}
Here, the subscript $B$ denotes the block matrix operation. 
Following this block extraction, the matrix $(\bm{I} - \omega^k \bm{d}^{\pm 1})$ becomes invertible, enabling us to derive the result above. 
It is evident that the block matrix $\hat{H}_B$ satisfies the chiral symmetry $P^{-1}\hat{H}_BP =-\hat{H}_B^T$, where $P$  denotes the backward identity matrix.
Consequently, this matrix exhibits spectral antisymmetry \cite{altland1997}.
 Given its odd dimension, it must have a zero eigenvalue, implying the existence of a nontrivial solution for this subsystem. 
 Furthermore, the block matrix is guaranteed to be nonzero in the gray- and red-labeled regions.
Using a recurrence relation, one can show that this matrix’s nontrivial solution is unique, with all components $\bm{\tilde{X}}_B$ nonzero. 
After solving the two subsystems, an additional constraint relating them remains. 
For the case in Figure \ref{figchessboard}, there is a linear constraint involving 
 $x_8$, $x_9$, and $x_{10}$. 
This means two linearly independent solutions exist for the full matrix $\hat{H}$ in Equation (\ref{eqEQ1}).
Reversing our reasoning, the matrix $\bm{T}\bm{S}_0$  in Equation (\ref{eqKYBE}) has eigenvalues $\omega^k$ for $k = 0, 1, \cdots, N-1$, 
with each eigenspace two-dimensional.
 In other words, this matrix is diagonalizable with eigenvalues $\omega^k$. 
 Therefore, we can conclude that Equation (\ref{eqKYBE}) indeed holds.

\section{Algebraic structures underlying the Kaleidoscope Yang-Baxter Equation}

We rewrite the Kaleidoscope Yang-Baxter Equation in a more concise form. For even $N$, we rewrite the following two matrices:

\begin{align}
\bm{T}:=
\begin{bmatrix}
\bm{t} & \bm{0} \\
\bm{0} & \bm{t}^{-1}  \\
\end{bmatrix},
\quad
\bm{\Lambda}:=
\begin{bmatrix}
\bm{s} & \bm{s}  \\
-\bm{s}  & -\bm{s}   \\
\end{bmatrix}.
\label{eqTLambda}
\end{align}
Here, $\bm{t}$ and $\bm{s}$ are defined as in Equation (\ref{eqts}). Notice that the dependence on $\theta$ is encoded in the $\bm{S}$ matrix. Thus, Equation (\ref{eqKYBE}) takes the following form:
\begin{equation}
    \left[\bm{T}(\bm{I} + z\bm{\Lambda})    \right]^N = \bm{I}.
    \label{eqKYBE2}
\end{equation}
Expanding the left-hand side of Equation (\ref{eqKYBE2}) in powers of $z$ yields a series of identities. For example, $(\bm{T}\bm{\Lambda})^N = \bm{0}$. Furthermore, we find that the $\bm{T}$ and $\bm{\Lambda}$ matrices satisfy the following properties:
\begin{enumerate}
    \item $\bm{T}^N = \bm{I};$
    \item $\bm{\Lambda}^2 = \bm{0};$
    \item Conjecture: $\bm{\Lambda}\bm{T}^{k_1}\bm{\Lambda}\bm{T}^{k_2} \cdots \bm{T}^{k_M}\bm{\Lambda} = \bm{0}, $ for $M \geq \frac{N}{2}$ and $k_1, k_2, \cdots, k_M \in \mathbb{Z} $. 
\end{enumerate}
The third point above is a conjecture we have proposed, which is a generalization of a corollary of the Kaleidoscope Yang-Baxter Equation. 
So far, we have not been able to prove it.
We have only verified its validity for $N \leq 10$. If it holds, it provides some structural information about the Kaleidoscope Yang-Baxter Equation. 
We note that $\bm{T}$ and $\bm{\Lambda}$, as generators, form a monomial algebra $\mathcal{A}$ of Loewy-length $N/2$. This algebra can be decomposed into a semisimple part $\mathcal{S} = \mathrm{span}(\bm{T}^k), k \in \mathbb{Z}$ and a nilpotent part  (Jacobson root)  $\mathcal{J} = \langle \bm{\Lambda} \rangle$:
\begin{equation}
    \mathcal{A} = \mathcal{S} + \mathcal{J}.
\end{equation}
We note that for the quantum mechanical scattering problems under discussion, the quantum contribution appears in the matrix $\bm{\Lambda}$. 
For  classical scattering, there is no need to introduce $\bm{\Lambda}$--it suffices to consider $\bm{t}$ in (\ref{eqts}) alone. 
The semisimple part of this algebra is generated by $\bm{T}$, which indicates that, when we use the Bethe ansatz to discuss scattering problems.
{\it The quantum mechanical scattering can be understood as a torsion added to the classical scattering.}

To further explore the hidden mathematical structures in the the Kaleidoscope Yang-Baxter Equation, we apply the following similarity transformation to Equation (\ref{eqKYBE2}).
Let us introduce 
    \begin{equation}
        \bm{P}:=\begin{bmatrix}
            \bm{I} & \bm{I}  \\
            -\bm{I}  & \bm{0}   \\
        \end{bmatrix}, \quad
        \bm{P}^{-1}:=\begin{bmatrix}
            \bm{0} & -\bm{I}  \\
            \bm{I}  & \bm{I}   \\
        \end{bmatrix}
    \end{equation}
     with the similarity transformation, we have
    \begin{equation}
        \bm{P}^{-1}\bm{T}^p\bm{P}=
        \begin{bmatrix}
            \bm{t}^{-p} & \bm{0}  \\
            \bm{t}^{p}-\bm{t}^{-p}  & \bm{t}^{-p}   \\
        \end{bmatrix},
        \quad
        \bm{P}^{-1}\bm{\Lambda}\bm{P}=
        \begin{bmatrix}
            \bm{0} & \bm{s}  \\
            \bm{0}  & \bm{0}   \\
        \end{bmatrix}.
    \end{equation}
The above transformation is very convenient for our computations when expanding the Kaleidoscope Yang-Baxter Equation.
Note that $\bm{s}$ can be expressed as $\bm{s} = (u\bm{d} - u^{-1}\bm{d}^{-1})^{-1}$, with free parameter $u =\mathrm{e}^{\mathrm{i}\theta}$.
 This  means that $\bm{s}$ itself is a function of $\bm{d}$. In Equation (\ref{eqtd}), $\bm{t}$ and $\bm{d}$ serve as the generators of the quantum torus algebra, satisfying 
\begin{equation}
\bm{t}^N = \bm{d}^N = \bm{I}, \quad \bm{t}\bm{d} = \mathrm{e}^{2\pi\mathrm{i}/N} \bm{d}\bm{t}. 
\end{equation}
This mathematical structure was first noticed as a special sub-algebra in the study of quantum groups \cite{levendorskii1991}.
It is also widely presented in various branches of modern physics, including noncommutative field theory
\cite{douglas2001}, the quantum Hall effect \cite{bellissard1994}, and noncommutative geometry\cite{chatzistavrakidis2012}.

The  Kaleidoscope Yang-Baxter Equation yields a series of identities for the quantum torus algebra.
For example, expanding the Kaleidoscope Yang-Baxter Equation and examining each order in $z$, we obtain  the following identities  which hold for any $u \in \mathbb{C}$
\begin{align}
    \sum_{p_1+\cdots +p_M=N-1}\bm{t}^{\pm p_1}(u\bm{d}-u^{-1}\bm{d}^{-1})^{-1}
    (\bm{t}^{p_2}-\bm{t}^{-p_2})\cdots  (\bm{t}^{p_{M-1}}-\bm{t}^{-p_{M-1}})(u\bm{d}-u^{-1}\bm{d}^{-1})^{-1}\bm{t}^{\pm p_M}=0
\end{align}
for $p_1, \cdots, p_M \geq 1$.  These identities themselves are not easy to prove directly.
However, as consequences of the Kaleidoscope Yang-Baxter Equation, we are able to obtain this series of intricate results. 
This serves as further evidence of the nontrivial nature of the Kaleidoscope Yang-Baxter Equation.

\section{Conclusion}

In this work, we have thoroughly studied the generalization of integrability conditions proposed in Ref. [1] and established a systematic method for constructing Bethe ansatz solutions for a class of two-dimensional models.
We have reanalyzed  the LQZC model, derived its complete energy spectrum, and show why coordinate Bethe ansatz fails for symmetry groups $D_{2N} $ with $ N \geq 4$.
Using a complete decomposition method, we  have solved  Gaudin's Kaleidoscope model.
Notably, through numerical and analytical analysis of its exact solution, we have observed that Bethe ansatz solvability depends not only on the model's symmetry but also on the symmetry subspace where solutions are sought.
Furthermore, we have also obtained  a nontrivial self-consistency condition for the scattering matrix in term of the Kaleidoscope Yang-Baxter Equation, which exhibits rich mathematical structure and merits further investigation.
In future work, we will study this equation in detail and analyze its mathematical structures.
 We will also apply our developed methods to explore connections between integrability, quantum chaos, and topology.  To facilitate reproducibility, the full implementation and all scripts used to produce the results in this work are publicly available at https://github.com/qiuwenjie24/CodeSciPostFig-KYBT.

\section*{Acknowledgements}
We thank  Yunbo Zhang, Xiangguo Yin, Yangyang Chen and Song Cheng for insightful discussions.


\paragraph{Funding information}
X.W.G. and Y.C.Y. acknowledge support from the NSFC key grants No. 12134015 and  No. 92365202,  and the Innovation Program for Quantum Science and Technology Projects under  No. 2021ZD0302000 and No. 2023ZD0300404, and  the National Key R\&D Program of China under Grant No. 2022YFA1404104.  Y.C. Y is also supported by the National Natural Science Foundation of China under Grants No. 12274419 and the CAS Project for Young Scientists in Basic Research under Grant No. YSBR-055.

\section*{Appendix}
\begin{appendix}
\numberwithin{equation}{section}

\section{The Finite Element Method for $\delta$-interacting Problems} 
The finite element method (FEM), also known as finite element analysis (FEA), is a powerful and versatile numerical technique for solving a wide range of field problems governed by partial differential equations (PDEs) with given boundary conditions. These problems arise in diverse disciplines such as structural mechanics, heat transfer, fluid dynamics, mass transport, and electromagnetics \cite{pepper2005, brenner2008}. The core idea of FEM is to discretize a complex domain into smaller, simpler subdomains called finite elements. Within each element, the unknown field variables are approximated by simple functions, and local equations are derived. These local equations are then assembled into a global system that approximates the behavior of the entire domain, reducing the original PDE problem to a system of algebraic equations that can be efficiently solved using modern computational techniques \cite{hughes2012}, further developments were given in \cite{turner1956, argyris1954, babuska1972}. Today,  major engineering industries employ the FEM for virtual prototyping, design optimization, and performance assessment of complex products \cite{brenner2008, cook2007}. The mathematical rigor and computational efficiency of the FEM have led to its widespread adoption and continuous development.

In our work, the problem under consideration is a typical eigenvalue problem with $\delta$-function potential. 
 Rigorously speaking, we deal  with contact interaction potentials, which can also be interpreted as boundary conditions connecting the field across different regions. 
 To the best of our knowledge, neither such potentials nor the associated boundary conditions have been discussed in previous FEM studies. 
 Apart from this particular aspect, we also show that this  problem can be addressed in a very concise and efficient manner  by using the FEM.

The purpose of this appendix is to demonstrate  the application of FEM to the $\delta$-function potential  problems. 
We will provide a brief introduction to the FEM, with a particular focus on deriving the formulation of the matrices in the presence of contact interactions.
 The following study  is  essential for carrying out the computations presented  in the main text.

\subsection{The mesh and basis in FEM }
\begin{figure}[h]       
  \centering               %
  \includegraphics[width=0.95\textwidth]{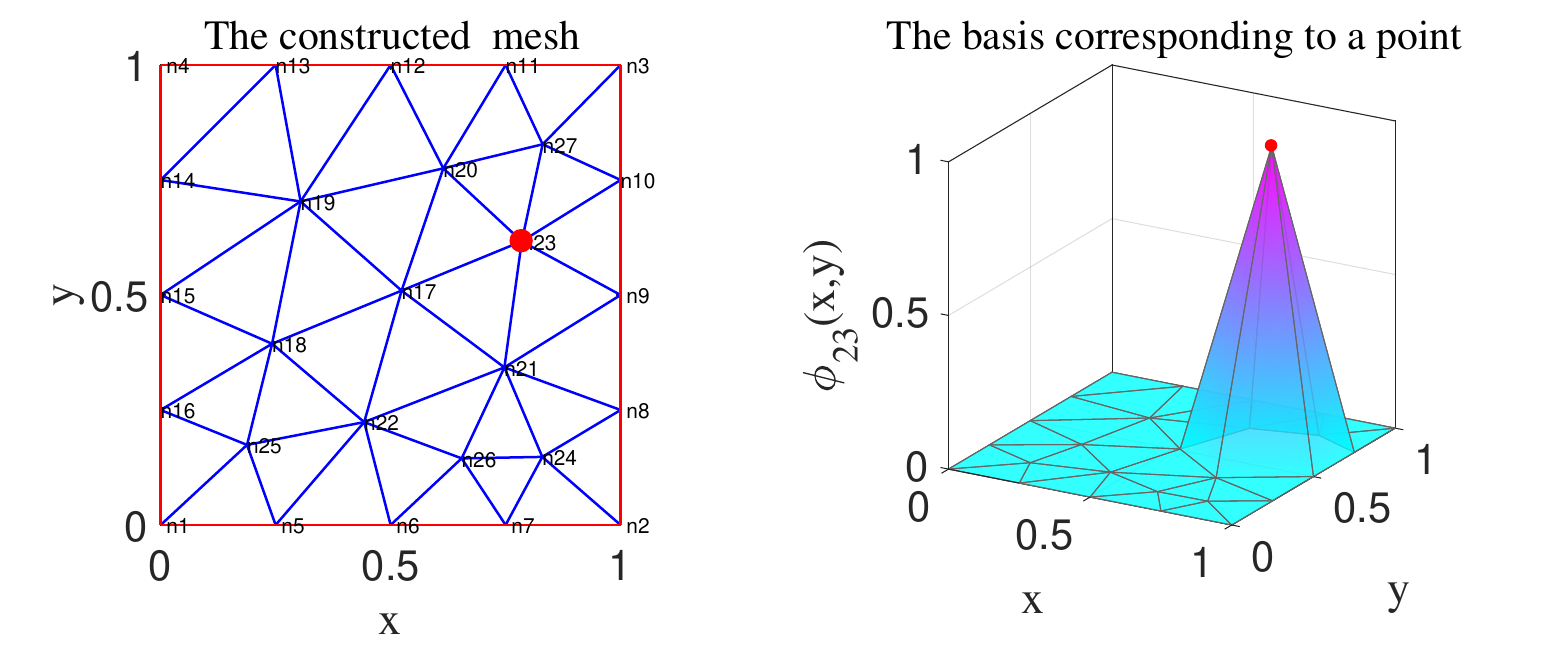} 
  \caption{The left panel illustrates the triangular partition, while the right panel shows a basis function corresponding to the point marked by the red dot (labeled 23) in the left panel.}
  \label{figFEM}       %
\end{figure}

The most crucial step in the finite element method is the partitioning of the solution domain, typically into triangular elements. Based on this partition, the solution space can be reduced to a finite-dimensional subspace whose basis functions are defined as follows: for each grid point, the corresponding basis function takes the value 1 at that point and 0 at all other grid points, and is linear within each triangle. As shown in Figure \ref{figFEM}, the left panel displays a triangular partition of the solution domain (here, the unit square). For each point in this partition, a basis function can be defined, which takes the form of a polyhedral cone in three-dimensional space. We denote this subspace $F$-space, and its basis are $\phi_n(x)$.

We continue to use the standard definition of the inner product in the Hilbert space of square-integrable functions ($L^2$-space), i.e., $\langle f, g \rangle := \int_\Omega \mathrm{d} \vec{x} f^*(\vec{x})g(\vec{x})$. For a given function $\vert f \rangle$ in the $L^2$-space, our goal is to find a point in the $F$-space that is as close to it as possible. This is equivalent to solving the minimization problem for the following loss function $\mathcal{L}$:
\begin{equation}
    \mathcal{L}=\vert\vert \, \vert  f\rangle - \sum_n b_n \vert \phi_n \rangle \, \vert \vert ^2
    = \langle f \vert f \rangle - \sum_n \left(b_n^* \langle \phi_n \vert f \rangle + b_n \langle f \vert \phi_n \rangle \right) + \sum_{n^\prime n} b_{n^\prime}^*b_n \langle \phi_{n^\prime}\vert \phi_n \rangle .
    \label{eqLossFunction}
\end{equation}
It is a typical quadratic optimization problem. According to $\frac{\partial \mathcal{L}}{\partial b_{n^\prime}^*} = 0$, we obtain the solution for the expansion coefficients $\bm{b}:=[b_1,b_2, \cdots,b_N]^T$ as 
\begin{equation}
    \bm{b} = \bm{M}^{-1} \bm{f},
\end{equation}
here, the components of the column vector $\bm{f}$ are the overlap integrals between $\vert f \rangle$ and the bases functions, i.e., $f_n = \langle \phi_n\vert f\rangle$, and the $\bm{M}$ is known as the ``mass matrix'' defined by 
\begin{equation}
    {M}_{n^\prime n} = \langle \phi_{n^\prime} \vert \phi_n \rangle .
    \label{eqMm}
\end{equation}
An important feature of the $\bm{M}$ matrix is its sparsity, which arises because its matrix elements $M_{n^\prime n}$ are nonzero if and only if the points labeled by $n$ and $n^\prime$ belong to same triangle. This sparsity is a key factor enabling the high efficiency of the FEM algorithm.

\begin{figure}[h]       
  \centering               %
  \includegraphics[width=0.95\textwidth]{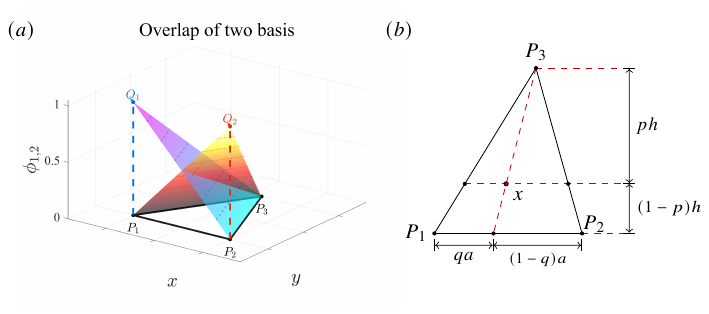} 
  \caption{Schematic for calculating the overlap integrals of basis functions. We focus on a small element, namely the triangle $\triangle P_1P_2P_3$. Here the basis function corresponding to $P_1$ is illustrated by the triangle $\triangle Q_1P_2P_3$ in the left panel, while the basis function corresponding to $P_2$ is illustrated by the triangle $\triangle P_1Q_2P_3$. The right panel illustrates the reparameterization of points on the triangle $\triangle P_1P_2P_3$. }
  \label{figFEM2}       %
\end{figure}

In the following, we provide a brief explanation of the computation of the mass matrix. According to the basic principles of the finite element method, it suffices to focus on calculating the overlaps of the basis functions within each small triangle, referring to Figure \ref{figFEM2}(a), we now compute the overlap integral between the basis functions $\phi_1$ and $\phi_2$, which correspond to $\triangle Q_1P_2P_3$ and $\triangle P_1Q_2P_3$, respectively, on the triangle $\triangle P_1P_2P_3$. To this end, we reparameterize the triangle according to the method shown in Figure \ref{figFEM2}(b). In this coordinate system, the area element is given by $\mathrm{d}S = ahq\mathrm{d}p\mathrm{d}q$, where $a$ and $h$ denote the base and the height of the triangle, respectively. It is straight forward to obtain
\begin{align}
     \int_{\triangle P_1P_2P_3} \mathrm{d} S \, \vert \phi_1(\vec{x}) \vert^2 &= \int_0^1\mathrm{d}p\int_0^1\mathrm{d}q \, ahp^2q=\frac{1}{3}S_{\triangle P_1P_2P_3}, \notag \\
    \int_{\triangle P_1P_2P_3} \mathrm{d} S \,\phi^*_1(\vec{x})\phi_2(\vec{x}) &= \int_0^1\mathrm{d}p\int_0^1\mathrm{d}q \, ahp(1-p)q=\frac{1}{6}S_{\triangle P_1P_2P_3},
    \label{eqOneElement}
\end{align}
here $S_{\triangle P_1P_2P_3} = \frac{1}{2}ah$ is simply the area of the triangle. Thus, to obtain the mass matrix (\ref{eqMm}), all we need to do is loop over each triangle and, for each triangle, compute and add its contribution to the $\bm{M}$ matrix according to the rules in (\ref{eqOneElement}). This process is referred to as "assemble" in FEM. After assembling, what we obtain is a set of sparse matrices, and the physical problem to be solved is reduced to a linear algebra problem involving these sparse matrices.

\subsection{The kinetic matrix, potential matrix and the eigenvalue problem}
 We consider the result of an operator $\hat{H}$ acting on a field (wave function) $\vert \psi \rangle$ in the $F$-space described by a column matrix $\bm{b}$ as $\vert \psi^\prime \rangle = \hat{H} \vert \psi \rangle = \sum_n b_n  \hat{H} \vert\phi_n \rangle$. Our goal is to find a wave function in the $F$-space, described by column matrix $\bm{b}^\prime$, such that it approximates $\vert \psi^\prime \rangle $ as closely as possible. This problem is entirely analogous to the quadratic optimization problem in (\ref{eqLossFunction}). Here we simply present the result:
 \begin{equation}
     \bm{b}^\prime = \bm{M}^{-1} \bm{H} \bm{b}. 
     \label{eqOperation}     
 \end{equation}
Equation (\ref{eqOperation}) completely characterizes the action of an operator in the $F$-space, where $\bm{M}$ is the mass matrix defined in (\ref{eqMm}) and the $\bm{H}$ matrix is defined similarly as 
\begin{equation}
    H_{n^\prime n} = \langle \phi_{n^\prime} \vert \hat{H} \vert \phi_n \rangle =  \int \mathrm{d} \vec{x} \, \phi_{n^\prime}^*(\vec{x}) \hat{H} \phi_n(\vec{x}).
    \label{eqHm}
\end{equation}

\subsubsection{The eigen problem}
In accordance with (\ref{eqOperation}), in the computational space of finite elements, the Hamiltonian $H$ is transformed into a matrix $\bm{M}^{-1}\bm{H}$, and thus the eigenvalue problem $\hat{H} \vert \psi \rangle = \epsilon \vert \psi \rangle $ is reduced to a generalized eigenvalue problem:
\begin{equation}
    \bm{H}\bm{b} = \epsilon \bm{M}\bm{b},
    \label{eqEigen}
\end{equation}
the $\epsilon$ and $\bm{b}$ in (\ref{eqEigen}) are the eigenvalues and eigenvectors that we are interested in.

In a broad class of physical problems discussed using the FEM (including the problem considered by this work), $\hat{H}$ consists of a kinetic part and a potential part $\hat{H} = \hat{T} + \hat{V}$, with the kinetic part given by the Laplacian operator $\hat{K}:= -\partial^2$ \cite{pepper2005, brenner2008, hughes2012}. Thus, the $\bm{H}$ matrix is also separated into the kinetic matrix $\bm{K}$ and the potential matrix $\bm{V}$:
\begin{equation}
    \bm{H} = \bm{K}+\bm{V}.
    \label{eqHeqKsumV}
\end{equation}
Now the FEM procedure reduces to two main tasks: (1) assembling the matrices $K_{n^\prime n} = \langle \phi_{n^\prime} \vert \hat{K} \vert \phi_n \rangle$ and $V_{n^\prime n} = \langle \phi_{n^\prime} \vert \hat{V} \vert \phi_n \rangle$; and (2) solving the generalized eigenvalue problem in (\ref{eqEigen}). For (2), there already exist many well-established algorithms and software packages capable of performing this task, so it is unnecessary for us to consider the algorithmic details. Therefore, in the following, we will focus on the problem of assembling kinetic and potential matrices.

\subsubsection{The kinetic matrix}
\begin{figure}[h]       
  \centering               
  \includegraphics[width=0.9\textwidth]{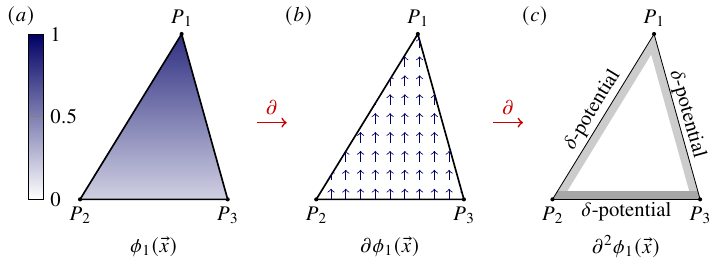} 
  \caption{(a) The basis function $\phi_1$: it is linear, increasing from $0$ to $1$ along the height of the triangle;
(2) The gradient of the basis function $\phi_1$: it is a constant vector field in the direction of the height;
(3) The divergence of the gradient of the basis function, i.e., $\partial^2 \phi_1$: it is a sum of $\delta$-functions localized on the edges of the triangle, each with a different coefficient.}
  \label{figFEM3}       
\end{figure}

For the calculation of the kinetic energy matrix, the key point is the action of the derivative operator $\partial$ on the basis functions. As shown in Figure (\ref{figFEM3}), we consider the basis function $\phi_1$ defined on the region of triangle $\triangle P_1P_2P_3$. This function increases linearly in the direction perpendicular to the base $P_2P_3$ of the triangle, as illustrated in Figure \ref{figFEM3}(a). A direct consequence of this linearity is that $\partial \phi_1$ results in a constant vector field which is perpendicular to the base with magnitude $\frac{1}{h_1}$ ($h_1$ is the height of the triangle with respect to the base $P_2P_3$), see Figure \ref{figFEM3}(b). It is evident that this gradient field is discontinuous across each of the three edges of the triangle, and thus taking the divergence of this gradient field yields three $\delta$-potential functions supported on the three edges of the triangle, see Figure \ref{figFEM3}(c). We will compute their coefficients in the following.

\begin{figure}[h]       
  \centering               
  \includegraphics[width=0.7\textwidth]{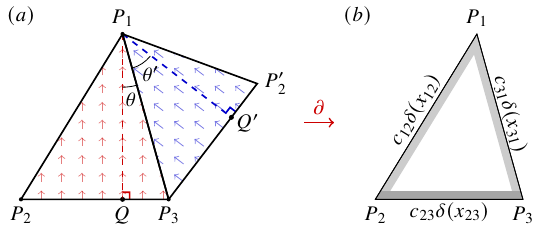} 
  \caption{(a) The discontinuity of the gradient of $\phi_1$ across two adjacent triangular regions;
(b) This gradient discontinuity determines that the result of $\partial^2 \phi_1$ is a set of $\delta$-functions localized on the boundaries of the triangle. Here, we consider only the contribution from a single region.}
  \label{figFEM4}       
\end{figure}

Consider the behavior of the basis function $\phi_1$ corresponding to $P_1$ in two adjacent regions $\triangle P_1P_2P_3$ and $\triangle P_1P_3P_2^\prime$. Please refer to Figure \ref{figFEM4}. It is clear that the gradient of the basis function is discontinuous across the two regions; however, along the interface line $P_1P_3$, the projections of the gradients onto the direction along the line $P_1P_3$ are continuous on both sides. In Figure \ref{figFEM4}(a), we have drawn the heights $P_1Q$ and $P_1Q^\prime$ of the two triangles, respectively.  Then the gradients of $\phi_1$ at the two regions can be evaluated:
\begin{equation}
    \nabla \phi_1 \big|_{\vec{x} \in \triangle P_1P_2P_3} = \vert QP_1 \vert^{-1} \bm{\hat{r}}_{QP_1}, \quad
     \nabla \phi_1 \big|_{\vec{x} \in \triangle P_1P_3P_2^\prime} = \vert Q^\prime P_1 \vert^{-1} \bm{\hat{r}}_{Q^\prime P_1}.
     \label{eqGradientPhi1}
\end{equation}
here $\bm{\hat{r}}_{QP_1}$ and $\bm{\hat{r}}_{Q^\prime P_1}$ are the unit vectors in the direction of $\overrightarrow{QP_1}$ and $\overrightarrow{Q^\prime P_1}$ respectively. In the direction parallel to $\overrightarrow{P_3P_1}$, the two gradients are equal, which is evident from 
\begin{equation}
\bm{\hat{r}}_{P_3P_1} \cdot\nabla \phi_1 \big|_{\vec{x} \in \triangle P_1P_2P_3} = \bm{\hat{r}}_{P_3P_1} \cdot\nabla \phi_1 \big|_{\vec{x} \in \triangle P_1P_3P_2^\prime} = \vert P_3 P_1 \vert ^{-1}.
\end{equation}
Therefore, at the interface between the two regions, the discontinuity of the gradient of $\phi_1$ across the two sides exists only in the direction perpendicular to the interface. This is also why $\partial^2 \phi_1$ results in a $\delta$-function localized on the interface. We can write the expression for $\partial^2 \phi_1$ in a single triangular region as follows (see Figure \ref{figFEM4}(b)):
\begin{equation}
    \partial^2 \phi_1 (\vec{x}) = c_{23} \delta(x_{23}(\vec{x})) 
    + c_{31}\delta(x_{31}(\vec{x})) + c_{12}\delta(x_{12}(\vec{x})),
    \label{eqPartial2}
\end{equation}
the functions $x_{23}(\vec{x}), x_{31}(\vec{x})$ and $x_{12}(\vec{x})$ denote the distance from $\vec{x}$ to the edge $P_2P_3, P_3P_1$ and $P_1P_2$ respectively. 

The coefficient in front of the $\delta$-function in (\ref{eqPartial2}) is determined by the discontinuity of the two gradient fields. For example, on $P_1P_3$ we can compute:
\begin{equation}
    \partial^2 \phi_1 \big|_{\vec{x} \in P_3P_1} = \frac{\tan \theta + \tan \theta^\prime}{\vert P_3 P_1 \vert}\cdot \delta(x_{31}(\vec{x})),
    \label{eqPartial2Phi1}
\end{equation}
here $\theta:=\angle P_3P_1Q$, $\theta^\prime := \angle P_3P_1Q^\prime$,  see Figure \ref{figFEM4}(a). Note that the coefficients of the $\delta$-functions in the above expression (\ref{eqPartial2Phi1}) are contributed separately by the two triangular regions. Since we will eventually loop over all triangles, it suffices to consider a single triangle. Figure \ref{figFEM4}(b) illustrates the case where only a single triangle is considered, and the contribution to the $\delta$-function arises solely from this triangle. The calculation in Equation (\ref{eqPartial2Phi1}) has already provided the coefficient in front of $\delta(x_{31}(\vec{x})$, and the other two coefficients can likewise be obtained directly by computing the second derivative of $\phi_1$. After some straightforward manipulations (involving only elementary plane geometry), the results for this set of coefficients in Equation (\ref{eqPartial2}) and Figure \ref{figFEM4}(b) can be written as follows:
\begin{equation}
    c_{12}=\frac{\bm{\hat{r}}_{P_1P_2}\cdot\overrightarrow{P_2P_3}}{2S_{\triangle P_1P_2P_3}}, \quad
    c_{23}=\frac{\bm{\hat{r}}_{P_2P_3}\cdot\overrightarrow{P_2P_3}}{2S_{\triangle P_1P_2P_3}}, \quad
    c_{31}=\frac{\bm{\hat{r}}_{P_3P_1}\cdot\overrightarrow{P_2P_3}}{2S_{\triangle P_1P_2P_3}}. 
    \label{eqCoefficient}
\end{equation}
Our current task is to compute the overlap between $\phi_1, \phi_2, \phi_3$ and the field $\partial^2 \phi_1$ shown in Figure \ref{figFEM4}(b). This calculation is straightforward: due to the presence of the $\delta$-function, we only need to evaluate a one-dimensional integral along each edge of the triangle. Substituting (\ref{eqPartial2}) and (\ref{eqCoefficient}) into the definition, it is evident that we can obtain:
\begin{align}
     \langle \phi_1 \vert \hat{K} \vert \phi_1 \rangle \big|_{\triangle P_1P_2P_3} &= -\int_{\triangle P_1P_2P_3} \mathrm{d} S \, \phi^*_1(\vec{x}) \partial^2 \phi_1(\vec{x}) = -\frac{\overrightarrow{P_2P_3}\cdot \overrightarrow{P_2P_3}}{4S_{\triangle P_1P_2P_3}} ,\notag \\
     \langle \phi_2 \vert \hat{K} \vert \phi_1 \rangle \big|_{\triangle P_1P_2P_3} &= -\int_{\triangle P_1P_2P_3} \mathrm{d} S \, \phi^*_2(\vec{x}) \partial^2 \phi_1(\vec{x}) = -\frac{\overrightarrow{P_2P_3}\cdot \overrightarrow{P_3P_1}}{4S_{\triangle P_1P_2P_3}} ,\notag \\
     \langle \phi_3 \vert \hat{K} \vert \phi_1 \rangle \big|_{\triangle P_1P_2P_3} &= -\int_{\triangle P_1P_2P_3} \mathrm{d} S \, \phi^*_3(\vec{x}) \partial^2 \phi_1(\vec{x}) = -\frac{\overrightarrow{P_2P_3}\cdot \overrightarrow{P_1P_2}}{4S_{\triangle P_1P_2P_3}} .
    \label{eqOneElementK}
\end{align}
The result of Equation (\ref{eqOneElementK}) is sufficient for assembling the kinetic energy matrix. We can simply loop over all triangles, compute the pairwise matrix elements on each triangle, and sum them up, in complete analogy with the computation for the mass matrix (\ref{eqMm}) and (\ref{eqOneElement}).

\subsubsection{The $\delta$-potential matrix}

\begin{figure}[h]       
  \centering               
  \includegraphics[width=0.82\textwidth]{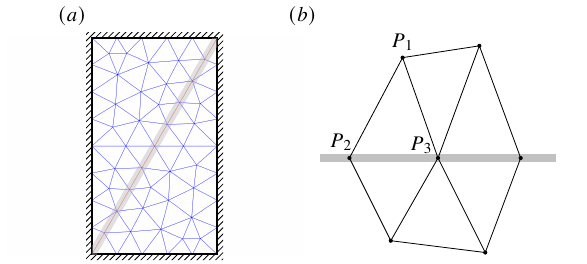} 
  \caption{(a) A convenient way to generate the mesh is to ensure that both the boundaries and the $\delta$-potential barriers are located along the edges of the small triangles. The gray lines represent the $\delta$-function barriers.
  (b) Only those triangles that have at least one edge lying on the barrier need to be considered in the calculation; for example, the triangle $\triangle P_1P_2P_3$. }
  \label{figFEM5}       
\end{figure}

Compared to the calculation of the kinetic energy matrix, the computation of the potential energy matrix is much simpler: we only need to evaluate the overlap integrals between the basis functions, weighted by the potential. In particular, for our problem, this calculation can be further simplified. Since we have the freedom to generate the grid points, we can ensure that every point on the $\delta$-potential barrier lies on the edge of a triangle (see Figure \ref{figFEM5}(a)), making the resulting potential matrix both easier to compute and even more sparse.

Suppose that the potential $\hat{V}$ is described by the $\delta$-function $V(\vec{x}) = c\delta(r(\vec{x}))$, here $c$ is the coupling constant, and $r(\vec{x})$ denotes the distance from $\vec{x}$ to the barrier. Then the matrix elements $\langle \phi_{n^\prime} \vert \hat{V} \vert \phi_n \rangle$ are nonzero if and only if the points corresponding to $n^\prime$ and $n$ both lie on the barrier, and they are either identical or adjacent. As before, we only need to consider each small triangle individually and then obtain the result by summing their contributions. Moreover, only those triangles that share an edge with the delta barrier contribute to the final result, such as the triangle $\triangle P_1P_2P_3$ in Figure \ref{figFEM5}(b). In this triangle, the overlap between the basis function corresponding to $P_1$ and the barrier is zero, so it can be neglected. Only the basis functions corresponding to $P_2$ and $P_3$ have nonzero overlap with the barrier, which reduces to one-dimensional linear functions on the segment $P_2P_3$. Thus, the calculation is straightforward, and we list the nonzero matrix elements as follows:
\begin{align}
\langle \phi_2 \vert \hat{V} \vert \phi_3 \rangle \big|_{\triangle P_1P_2P_3} &= \langle \phi_3 \vert \hat{V} \vert \phi_2 \rangle \big|_{\triangle P_1P_2P_3}=\frac{c}{2}\int_{\triangle P_1P_2P_3} \mathrm{d} S \, \phi^*_2(\vec{x}) \delta(x_{23}(\vec{x})) \phi_3(\vec{x}) = \frac{c}{12}\vert P_2P_3 \vert \notag \\
     \langle \phi_2 \vert \hat{V} \vert \phi_2 \rangle \big|_{\triangle P_1P_2P_3} &= \frac{c}{2}\int_{\triangle P_1P_2P_3} \mathrm{d} S \, \phi^*_2(\vec{x}) \delta(x_{23}(\vec{x})) \phi_2(\vec{x}) = \frac{c}{6}\vert P_2P_3 \vert \notag \\
     \langle \phi_3 \vert \hat{V} \vert \phi_3 \rangle \big|_{\triangle P_1P_2P_3} &= \frac{c}{2}\int_{\triangle P_1P_2P_3} \mathrm{d} S \, \phi^*_3(\vec{x}) \delta(x_{23}(\vec{x})) \phi_3(\vec{x}) = \frac{c}{6}\vert P_2P_3 \vert 
    \label{eqOneElementV}.
\end{align}
Note that in the above results (\ref{eqOneElementV}), $c$ denotes the coupling constant in front of the $\delta$-function, and we have included a factor of $\frac{1}{2}$ because the region we consider lies on one of the two total sides of the barrier. The result (\ref{eqOneElementV}) is sufficient for assembling the $\bm{V}$ matrix.

In summary, the results of Equations (\ref{eqOneElementK}) and (\ref{eqOneElementV}) allow us to assemble the matrix $\bm{H}$ according to Equation (\ref{eqHeqKsumV}), while the result for $\bm{M}$ is given in Equation (\ref{eqOneElement}). After completing these calculations, the problem is reduced to the generalized eigenvalue problem (\ref{eqEigen}). By applying existing numerical algorithms, we obtain an approximate solution to the original eigenvalue problem in the $F$-space. The above constitutes the entire computational procedure of the finite element method.

\subsection{Some remarks}
Please note that our discussion above is limited to the computations required for this work; in practice, the finite element method encompasses much more than what is presented in this appendix. Moreover, we have not addressed the problem of boundary conditions, which is a crucial aspect of FEM. However, in our current calculations, the boundary conditions are very simple—we consider only Dirichlet boundary conditions, where the wave function vanishes at the boundary. In this case, we simply remove from the $F$-space any basis functions corresponding to points on the boundary. Therefore, we will not elaborate on this further.

Our derivation differs slightly from the traditional approach \cite{pepper2005, brenner2008, hughes2012}. First, we have addressed the problem of the contact interaction potential, which is a distinctive feature of the model studied here and has not received much attention in previous FEM literature. The more significant difference lies in our treatment of the kinetic matrix. Drawing primarily on the modern perspective of  \cite{larson2010}, we combined this with techniques for handling $\delta$-functions to complete the calculation of the kinetic matrix. This approach avoids the use of the Green's integral formula and the associated discussion of weak solutions. From the viewpoint of someone familiar with Bethe ansatz, I believe this derivation is simpler and more direct.

\end{appendix}





\bibliography{KYBE.bib}


\end{document}